\documentclass[12pt,a4paper]{article}
 
\usepackage{lineno}
\usepackage{graphicx} 
\usepackage[margin=1in]{geometry}
\usepackage{amsmath}
\usepackage{amssymb}
\usepackage{graphicx}
\usepackage{float}
\usepackage{xcolor}
\usepackage[percent]{overpic}
\usepackage{tabularx}
\usepackage{array}
\usepackage{multirow}
\usepackage{threeparttablex}
\usepackage{caption} 
\usepackage{indentfirst}
\usepackage{sectsty}
\usepackage{titlesec} 
\titlespacing{\section}{0pt}{\parskip}{-\parskip} 
\titlespacing{\subsection}{0pt}{\parskip}{-\parskip} 
\titlespacing{\subsubsection}{0pt}{\parskip}{-\parskip} 
\sectionfont{\fontsize{12}{15}\selectfont}
\subsectionfont{\fontsize{12}{15}\selectfont}
\usepackage{tikz}
\usetikzlibrary{positioning,fit}
\usetikzlibrary{calc} 
\usetikzlibrary{shapes.geometric, arrows}
\usetikzlibrary{arrows.meta, positioning}
\tikzstyle{process} = [rectangle, minimum width=2.5cm, minimum height=1cm, text centered, draw=black]
\tikzstyle{arrow} = [thick,->,>=stealth] 
\renewcommand{\arraystretch}{1.3} 
\newcolumntype{L}[1]{>{\raggedright\arraybackslash}p{#1}} 
\newcolumntype{C}[1]{>{\centering\arraybackslash}p{#1}}  
\usepackage[skip=2pt]{caption} 
\usepackage{enumitem}
\usepackage{setspace} 
\usepackage{booktabs}
\usepackage{longtable}
\AtBeginDocument{%
    \setlength{\abovedisplayskip}{4pt}%
    \setlength{\belowdisplayskip}{4pt}%
    \setlength{\abovedisplayshortskip}{2pt}%
    \setlength{\belowdisplayshortskip}{2pt}%
}
\usepackage[super,sort&compress]{natbib}
\usepackage{url}

\usepackage{titling}
\usepackage{authblk}
\pretitle{\centering\Large\normalfont}
\posttitle{\par\vspace{-0.8em}}

\title{A Unified Framework to Prioritize Population and Precision Health Research}
\author[1,2]{Yuning Wang\thanks{Corresponding author: Yuning Wang, Email:yuningyn.wang@mail.utoronto.ca}}
\author[2,3]{Petros Pechlivanoglou}
\author[1,2,4]{Anna Heath}
\affil[1]{Division of Biostatistics, Dalla Lana School of Public Health, University of Toronto, Toronto, Ontario, Canada}
\affil[2]{Child Health Evaluative Sciences, The Hospital for Sick Children, Toronto, Ontario, Canada}
\affil[3]{Institute of Health, Policy, Management and Evaluation, University of Toronto, Toronto, Ontario, Canada}
\affil[4]{Department of Statistical Science, University College London, London, England}
\date{}

\begin{document}
{\onehalfspacing \maketitle}
\vspace{-3.5em}
\section*{Abstract}
{
\begingroup
\setstretch{1}
\setlength{\parindent}{0pt}
%\fontsize{10.5}{13.7}\selectfont
\textbf{Background}: In an era of precision health, researchers must often prioritize among competing research directions addressing different sources of uncertainty and variation. Future research may aim to reduce parameter uncertainty (e.g., improve the average treatment effect estimate) or explain individual-level heterogeneity arising from currently unknown characteristics (e.g., by identifying new biomarkers), known as exploring unexplained heterogeneity. Given limited research budgets, identifying the most valuable research direction is important. Existing frameworks evaluate these research directions separately, preventing researchers from prioritizing research. Thus, we develop a unified framework for prioritizing research by jointly estimating the value of resolving parameter uncertainty and unexplained heterogeneity. 

\textbf{Methods}: Our unified framework comprises six uncertainty states, defined by combinations of available parameter and unexplained heterogeneity information, extending concepts from the Value of Unexplained Heterogeneity framework. The value of a potential research direction is the difference in value between two states. We formalize an approach to define counterfactual outcomes across interventions to compute value and suggest ways to reduce the computational burden of value estimation. We illustrate the framework using a three-option microsimulation model. 

\textbf{Results}: We directly compare research value across different sources of uncertainty and variation. In the illustrative example, exploring unexplained heterogeneity had the greatest value. Comparing the value of resolving unexplained heterogeneity across individual-level outcome components (e.g., health states, costs, and side effects) indicated little value in personalizing interventions based on whether patients entered the sicker state.

\textbf{Conclusion}: Our framework enables direct, quantitative comparisons of the value of different research priorities and identifies the one with the greatest potential to improve population-level cost-effectiveness. It serves as a practical tool for performing research prioritization in the era of precision health.
\par
\endgroup
}

\clearpage
\section*{Highlights}
\begin{itemize}
    \item Prioritizing research investments in an era of precision health requires comparing the value of competing research directions: (1) identifying novel strategies for personalizing healthcare decisions, such as novel biomarkers that explain individual-level heterogeneity, and (2) improving population-level estimates of model parameters (e.g., average treatment effects). 
    \item We introduce a unified framework to compute the potential value of research that targets both parameter uncertainty and unexplained heterogeneity, enabling direct comparison and prioritization of competing research directions.
    \item We evaluate the potential value of research across a range of assumptions about the counterfactual outcomes for individuals receiving alternative interventions.
\end{itemize}

\section*{Introduction}
Advances in technology have led to growing recognition that individual-level heterogeneity (e.g., in molecular, radiographic, or physiologic markers) can impact disease development and progression. \cite{Fountzilas2022, Duan2024} This has driven a shift from conventional population-level "one-size-fits-all" interventions toward precision health care, which targets interventions based on these individual-level markers to improve and maintain individual-level health. \cite{Fu2020, Duan2024, Schork2025} With the rapid development of precision health, researchers and policymakers increasingly face diverse study settings and research objectives, ranging from discovering novel markers for personalized intervention to generating broader population-level evidence. \cite{Fountzilas2022, Duan2024} However, the costs of conducting research are high, and the resources and budgets for studies are generally limited. \cite{Mulcahy2025, Payne2023} Therefore, research prioritization is key. 

Previous studies have developed methods to estimate the value of research targeting personalized treatments, known as the Value of Unexplained Heterogeneity (VoUH) framework. \cite{Heath2022} VoUH methods calculate the additional value of providing patients with individualized or subgroup-optimal interventions by resolving, or partially resolving, \textit{unexplained heterogeneity}. \cite{Heath2022} Conceptually, in this framework, unexplained heterogeneity refers to the variation in patient outcomes that is currently not explained by observed characteristics. \cite{Heath2022} The VoUH framework therefore computes the potential value of research into novel personalized treatments. However, VoUH methods assume that parameters used in the estimation process (e.g., survival probabilities, quality of life, costs) are deterministic. In other words, parameter uncertainty is ignored. Importantly, parameter uncertainty can lead to suboptimal treatment decisions, which incur a cost to the healthcare system. \cite{Briggs2006} 

In contrast, other frameworks compute the value of resolving parameter uncertainty in decision-making. \cite{oostenbrink2008expected, wilson2015practical, heath2018calculating, jackson2021guide} Decision-analytic models (DAMs) synthesize the available evidence to compare the costs and health outcomes of healthcare interventions. \cite{bilcke2011accounting} Parameter uncertainty is incorporated into DAMs by assigning probability distributions to the model parameters. \cite{Briggs2006} Probabilistic analysis (PA) then propagates this uncertainty through the model to quantify its impact on model outputs and decision uncertainty. \cite{briggs2012model, fenwick2020value} Building on the PA, the Value of Information (VoI) framework quantifies the value of research to reduce parameter uncertainty. \cite{fenwick2020value, jackson2021guide} However, the VoI framework assumes that there is no heterogeneity between individuals, including both heterogeneity caused by observed characteristics (explained) and unexplained heterogeneity. The Value of Heterogeneity (VoH) framework was developed to estimate the value of understanding explainable heterogeneity. \cite{Espinoza2014} However, the VoH framework assumes complete information on individual characteristics related to outcome heterogeneity is available and unexplained heterogeneity is ignored. \cite{Espinoza2014} 

Thus, no framework can directly compare the value of research that targets parameter uncertainty to research that aims to detect novel markers to explore unexplained heterogeneity. This poses a challenge for prioritizing research in the era of precision health. To address this gap, we propose a novel framework that combines both parameter uncertainty and unexplained heterogeneity, building on the VoUH framework, to enable such comparisons and support research prioritization. 

In the following sections, we first introduce the sources of uncertainty and variation considered in the new framework. We then briefly review the VoI and VoUH frameworks to position our work as an extension of these frameworks. Next, we define our novel framework, which integrates parameter uncertainty and unexplained heterogeneity. We extend a key element of the VoUH framework, which requires the estimation of counterfactual health outcomes. Ways to improve computational efficiency of the estimation process are also suggested. We then discuss an illustrative example to demonstrate the application of our framework. We conclude by discussing the framework's limitations and suggesting future extensions.

\section*{Methods}
\subsection*{Decision Making With Multiple Sources of Uncertainty}
Healthcare decision making involves multiple sources of uncertainty and variation: stochastic (first-order) uncertainty, parameter (second-order) uncertainty, explainable heterogeneity, and structural uncertainty. \cite{Briggs2006, bojke2009characterizing, briggs2012model, Heath2023} In our framework, we focus on exploring the value of resolving stochastic and parameter uncertainty only. Here, stochastic uncertainty refers to the variations in individual-level outcomes after all observed individual characteristics and model parameters have been held constant, and is called \textit{unexplained heterogeneity} throughout the manuscript. \cite{Heath2022} Parameter uncertainty is uncertainty due to having limited available evidence to estimate the population-level model parameters. \cite{Briggs2006} The following section formalizes these two concepts.  

Throughout this study, we use the net benefit framework to evaluate healthcare outcomes, combining costs and effectiveness into a single measure. \cite{stinnett1998net, hoch2002something, Briggs2006} Specifically, we denote the individual-level net monetary benefit under intervention $d$ by $nb_d$:
\begin{equation} \label{eq:nb}
    nb_d = \lambda \times e_d - c_d,
\end{equation}
where $\lambda$ is the willingness-to-pay (WTP) threshold \cite{Briggs2006}, and $e_d$ and $c_d$ are the individual-level effectiveness and costs of intervention $d$, respectively, which could be calculated based on each patient's disease trajectory. We then use a vector $\boldsymbol{\mathcal{O}}$ to denote the elements of the individual-level disease trajectory, such as the health states patients experience, associated costs, and side effects. We call the elements of $\boldsymbol{\mathcal{O}}$ the outcome components, as they are combined to obtain the health economic outcomes $(c_d, e_d)$, and further compute $nb_d$. \cite{Heath2022} 

Although the true values of $\boldsymbol{\mathcal{O}}$ are not directly observable without monitoring the entire patient disease trajectory, their possible values can be simulated using individual-level DAMs, parameterized by $\boldsymbol{\theta}$, a vector of parameters that define the population-level behaviour of the model. \cite{petrou2011economic, bilcke2011accounting, PealozaRamos2015} Uncertainty in $\boldsymbol{\theta}$ is generally modeled as a high-dimensional joint probability distribution $p(\boldsymbol{\theta})$. \cite{baio2015probabilistic} When explainable and unexplained heterogeneity are ignored, population-average net benefits, conditional on $\boldsymbol{\theta}$, are computed by taking the conditional expectation of $nb_d$ with respect to $\boldsymbol{\mathcal{O}}$. This can be denoted as $\text{NB}_d(\boldsymbol{\theta})$ for intervention $d$:
\begin{equation*}
    \text{NB}_d(\boldsymbol{\theta}) = \mathbb{E}_{\boldsymbol{\mathcal{O}} \mid \boldsymbol{\theta}} \big[ nb_d (\boldsymbol{\theta}, \boldsymbol{\mathcal{O}} )\big] = \int f_d (\boldsymbol{\mathcal{O}}, \boldsymbol{\theta}) p_{\boldsymbol{\mathcal{O}} \mid \boldsymbol{\theta}}(\boldsymbol{\mathcal{O}} \mid \boldsymbol{\theta}) d \boldsymbol{\mathcal{O}}.
\end{equation*}
The expectation of $\text{NB}_d(\boldsymbol{\theta})$ is often estimated through PA, while $\text{NB}_d(\boldsymbol{\theta})$ itself is used to understand the impact of parameter uncertainty on the decision making process. Considering the complexity of $p(\boldsymbol{\theta})$, PA typically relies on Monte Carlo (MC) simulation.\cite{doubilet1985probabilistic, jackson2021guide} 

In individual-level DAMs, even for a given $\boldsymbol{\theta}$, patients with constant observed characteristics will still be simulated with differences in the individual-level outcome components $\boldsymbol{\mathcal{O}}$. \cite{siebert2003should, petrou2011economic, bilcke2011accounting} Thus, we define $p(\boldsymbol{\mathcal{O}} \mid \boldsymbol{\theta})$ as the distribution of the individual-level outcome components. This variability induced by this distribution differs from model parameter uncertainty $p(\boldsymbol{\theta})$ as it describes unexplained heterogeneity in individual outcomes. Therefore, the distribution of the individual-level net benefit $nb_d$ is defined by a joint distribution of $\mathcal{O}$ and $\boldsymbol{\theta}$:
\begin{equation*}
    nb_d = f_d(\boldsymbol{\mathcal{O}}, \boldsymbol{\theta}).
\end{equation*}
Thus, the individual-level net benefits simultaneously incorporate unexplained heterogeneity and parameter uncertainty. Both sources of uncertainty and variation could affect the choice of optimal intervention and, consequently, the potential outcomes of healthcare decisions. The VoI and VoUH frameworks were previously developed to quantify the value of future research that resolves parameter uncertainty and unexplained heterogeneity, respectively, but cannot currently combine these two sources of uncertainty and variation. \cite{jackson2021guide, Heath2022} However, as our proposed framework builds on them, we briefly review them here. 

\subsection*{Value of Information: Population-Level Research Reducing Parameter Uncertainty Only}
Based on the population-average net benefits NB$_d(\boldsymbol{\theta})$, the value of research aimed at reducing parameter uncertainty can be estimated using the VoI framework. Within this framework, the value of research varies with the level of parameter uncertainty reduction, from eliminating parameter uncertainty to reducing uncertainty for a subset of parameters or a single parameter. \cite{wilson2015practical, jackson2022value, Heath2023} The expected value of perfect information (EVPI) quantifies the maximum value of research targeting parameter uncertainty by comparing decisions under current evidence with those under perfect information: \cite{oostenbrink2008expected, thorn2016interpretation, heath2017review}
\begin{align} \label{eq:evpi}
    \text{EVPI} &= 
    \mathbb{E}_{\boldsymbol{\theta}} \left[ \max_d \mathbb{E}_{\boldsymbol{\mathcal{O}} \mid \boldsymbol{\theta}} \big[ nb_d (\boldsymbol{\mathcal{O}}, \boldsymbol{\theta}) \big] \right] - 
    \max_d \mathbb{E}_{\boldsymbol{\theta}} \Big[ \mathbb{E}_{\boldsymbol{\mathcal{O}} \mid \boldsymbol{\theta}} \big[ nb_d (\boldsymbol{\mathcal{O}}, \boldsymbol{\theta}) \big] \Big] \\ \notag
    &= \mathbb{E}_{\boldsymbol{\theta}} \left[ \max_d \text{NB}_d(\boldsymbol{\theta}) \right] - \max_d \mathbb{E}_{\boldsymbol{\theta}} \left[ \text{NB}_d(\boldsymbol{\theta}) \right].
\end{align} 
The first term computes the expected value of implementing the population-level optimal intervention with perfect parameter information, while the second term computes the expected value of using the current population-optimal intervention. 

The expected value of perfect partial information (EVPPI) estimates the value of eliminating uncertainty in specific parameter subsets and can compare how different parameters contribute to parameter uncertainty. \cite{brennan2007calculating, jackson2022value} Thus, EVPPI enables research prioritization across parameter sets. Finally, the expected value of sample information estimates the expected gains of collecting additional data through a specific future study. \cite{heath2018calculating} However, our proposed framework currently only incorporates EVPI, so we do not discuss these other VOI measures further. 

\subsection*{Value of Unexplained Heterogeneity: Personalized Treatment Research With Deterministic Parameters}
The VoUH framework calculates the value of research that discovers new evidence to personalize treatment, under the assumption of deterministic parameter values. \cite{Heath2022} The value of research within the VoUH framework also differs across levels of unexplained heterogeneity, ranging from resolving all unexplained heterogeneity to partially resolving unexplained heterogeneity in specific outcome components. \cite{Heath2022} Conditional on deterministic parameter values, the Maximum Value of Heterogeneity (MVoH) quantifies the value of resolving all unexplained heterogeneity:
\begin{equation} \label{eq:mvoh}
    \text{MVoH} = \mathbb{E}_{\boldsymbol{\mathcal{O}}} \left[ \max_d nb_d (\boldsymbol{\mathcal{O}} \mid \boldsymbol{\theta} )\right] - \max_d \mathbb{E}_{\boldsymbol{\mathcal{O}}} \left[ nb_d (\boldsymbol{\mathcal{O}} \mid \boldsymbol{\theta} ) \right]. 
\end{equation}
The first term computes the value of identifying individualized optimal interventions by resolving all unexplained heterogeneity, and the second term computes the value of providing a population-optimal intervention. 

The individual-level net benefit must be compared across treatments to identify individualized optimal treatment. Specifically, the optimal intervention depends on the relative ranking of the net benefit for each intervention for that individual, i.e., we must identify which intervention maximizes the net benefit for each individual. However, as individuals can receive only one intervention, $nb$ for alternative interventions is counterfactual. Although DAMs can simulate $nb$ for all interventions, they cannot establish the true ranking of interventions for each individual. Therefore, to compute the MVoH, the counterfactual must be specified explicitly. In the original VoUH framework, counterfactual outcomes were explored through scenario analyses using a correlation vector $\boldsymbol{\rho}$ and only models with two available interventions were considered. Critically, the MVoH strongly depends on $\boldsymbol{\rho}$. Positive values of $\boldsymbol{\rho}$ indicate that individuals who achieve relatively high (or low) net benefits under one intervention also tend to achieve relatively high (or low) net benefits under another intervention, reducing the value of personalization. Conversely, negative $\boldsymbol{\rho}$ values indicate that the relative performance of treatments varies more substantially across individuals, providing greater value of personalization. \cite{Heath2022} 

In the original VoUH framework, the MVoH was evaluated across values of $\boldsymbol{\rho}$ and if it remained high for most values, the Value of Perfect Outcome Prediction (VPOP) was suggested to prioritize research in personalized treatment strategies. \cite{Heath2022} Specifically, we assume a subset of outcome components $\boldsymbol{o}$ ($\boldsymbol{o}\subset\boldsymbol{\mathcal{O}}$), such as the occurrence of adverse events or cancer relapse, can be perfectly predicted, i.e., the unexplained heterogeneity in $\boldsymbol{o}$ is fully resolved. VPOP compares the contributions of unexplained heterogeneity underlying different $\boldsymbol{o}$ to overall unexplained heterogeneity. \cite{Heath2022} 
\begin{equation} \label{eq:vpop}
    \text{VPOP}_{\rho_{\boldsymbol{o}}} = \mathbb{E}_{\boldsymbol{o}} \left[ \max_d \Big\{ \mathbb{E}_{\boldsymbol{\mathcal{O}}^- \mid \boldsymbol{o}} \big[ nb_d (\boldsymbol{\mathcal{O}}^-, \boldsymbol{o} \mid \boldsymbol{\theta}) \big] \Big\}\mid \rho_{\boldsymbol{o}} \right] - \max_d \mathbb{E}_{\boldsymbol{\mathcal{O}}} \left[ nb_d (\boldsymbol{\mathcal{O}} \mid \boldsymbol{\theta}) \mid \rho_{\boldsymbol{o}} \right],
\end{equation}
where $\boldsymbol{\mathcal{O}}^-$ is the remaining outcomes, and $\rho_{\boldsymbol{o}}$ is the correlation between the expected individual-level net benefits with respect to $\boldsymbol{\mathcal{O}}^-$ across the two treatments. The first term in Eq.~\ref{eq:vpop} computes the value of providing patients with an optimal intervention conditional on the perfectly known $\boldsymbol{o}$. A higher VPOP indicates greater potential value of research and, therefore, a higher priority for resolving the unexplained heterogeneity in corresponding $\boldsymbol{o}$. 

\subsection*{Expected Value of Unexplained Heterogeneity: Research With Multiple Sources of Uncertainty and Variation} 
The VoI and VoUH frameworks provide evidence for prioritizing research, while ignoring unexplained heterogeneity or parameter uncertainty, respectively. However, researchers could target research towards reducing parameter uncertainty or identify novel markers for precision medicine. Comparable value of research across these varied research directions is therefore required. Thus, we propose a novel framework that takes into consideration both parameter uncertainty and unexplained heterogeneity: the Expected Value of Unexplained Heterogeneity (EVoUH) framework. 

In the EVoUH framework, \textit{parameter uncertainty} includes two levels, current and perfect information, where perfect parameter information means no remaining parameter uncertainty. Unexplained heterogeneity, which means changes to the \textit{optimal intervention}, has three levels, population, outcome-conditional, and individualized optimal intervention. Outcome-conditional optimal interventions are those based on perfectly predicting different outcome components $\boldsymbol{o}$. Table \ref{tab:framework} displays the framework where each cell represents a distinct level of information on parameter uncertainty and unexplained heterogeneity that is assumed to be available for decision making. These different ''states'' of research are each associated with a value that is computed using the individual-level net benefits $nb_d (\boldsymbol{\mathcal{O}}, \boldsymbol{\theta})$, which is a joint function of model parameters $\boldsymbol{\theta}$ and outcomes $\boldsymbol{\mathcal{O}}$. Transitions between states represent a potential research direction. By comparing the differences between states, the EVoUH framework informs research prioritization across diverse directions. 
{
\begin{table}[H]
    \centering 
    \renewcommand{\arraystretch}{1.0} 
    \begin{threeparttable} 
        {\centering 
            \begin{tabular}{|L{4.5cm}|C{4cm}|C{4cm}|} 
                \hline 
                \multirow{2}{*}[0pt]{\raggedright Optimal Intervention} & \multicolumn{2}{c|}{Statistical Parameter Information} \\ 
                \cline{2-3} 
                & Current & Perfect \\ 
                \hline 
                Population       & A & B \\ 
                \hline 
                Outcome-conditional$^*$     & C & D \\ 
                \hline 
                Individualized         & E & F \\ 
                \hline 
            \end{tabular} 
        } 
    \begin{tablenotes}[flushleft] 
        \footnotesize 
        \item \textit{$^*$}Patients are provided with their optimal interventions based on perfectly predicted outcome components. 
    \end{tablenotes} 
    \end{threeparttable}
    \caption{\textcolor{black}{The expected value of unexplained heterogeneity (EVoUH) Framework}} 
    \label{tab:framework} 
    \vspace{-0.7\baselineskip}
\end{table} 
}
In Table \ref{tab:framework}, $B - A$ computes the EVPI (Eq.~\ref{eq:evpi}), providing an upper bound for research resolving parameter uncertainty.  $E - A$ computes the  MVoH under current parameter uncertainty (MVoH$_{cur}$), which is the upper bound for research exploring new individualized intervention allocation strategies (e.g., novel biomarkers) under current parameter information:
\begin{equation} \label{eq:mvoh_cur}
    \text{MVoH}_{cur} = \mathbb{E}_{\boldsymbol{\mathcal{O}}} \Big[ \max_d \mathbb{E}_{\boldsymbol{\theta} \mid \boldsymbol{\mathcal{O}}} \big[ nb_d (\boldsymbol{\mathcal{O}}, \boldsymbol{\theta}) \big] \Big] - \max_d \mathbb{E}_{\boldsymbol{\mathcal{O}}} \Big[ \mathbb{E}_{\boldsymbol{\theta} \mid \boldsymbol{\mathcal{O}}} \big[ nb_d (\boldsymbol{\mathcal{O}}, \boldsymbol{\theta}) \big] \Big].
\end{equation}
The first term is the value of state $E$, where patients receive individualized optimal interventions, as heterogeneity is fully resolved, but parameter uncertainty is not reduced. 

In addition to this, the difference between states $A$ and $F$ represents the maximum achievable value for future research, where both parameter uncertainty and unexplained heterogeneity are resolved simultaneously. We define this maximum value as the Expected Maximum Value of Heterogeneity (EMVoH):
\begin{equation} \label{eq:emvoh}
    \text{EMVoH} = \mathbb{E}_{\boldsymbol{\mathcal{O}}} \Big[ \mathbb{E}_{\boldsymbol{\theta} \mid \boldsymbol{\mathcal{O}}} \big[ \max_d nb_d (\boldsymbol{\mathcal{O}}, \boldsymbol{\theta}) \big] \Big] - 
    \max_d \mathbb{E}_{\boldsymbol{\mathcal{O}}} \Big[ \mathbb{E}_{\boldsymbol{\theta} \mid \boldsymbol{\mathcal{O}}} \big[ nb_d (\boldsymbol{\mathcal{O}}, \boldsymbol{\theta}) \big] \Big].
\end{equation}
The first term is the value of $F$, an idealized state with perfect parameter and unexplained heterogeneity information. The second term is the value of $A$, the current state without additional information. As the maximum achievable value, EMVoH provides an upper bound for all feasible research. Thus, if EMVoH is associated with substantial value and research resolving parameter uncertainty and exploring unexplained heterogeneity are both possible, we suggest comparing EVPI and MVoH$_{cur}$ to prioritize a more valuable direction. Specifically, with limited research resources, if MVoH$_{cur}$ exceeds EVPI, prioritize research that resolves unexplained heterogeneity over studies that reduce parameter uncertainty. 

\subsection*{Expected Value of Perfect Outcome Prediction} 
When MVoH$_{cur}$ is higher, a natural follow-up question is which outcome component-related unexplained heterogeneity to explore first. As in the VoUH framework's move from MVoH to VPOP, the EVoUH framework continues to compare the value of research targeting different outcome components, where such research provides outcome-conditional optimal interventions. 

Assuming outcome components $\boldsymbol{o}$ can be perfectly predicted, the full set of outcome components $\boldsymbol{\mathcal{O}}$ is decomposed into $\boldsymbol{o}$ and the remaining components $\boldsymbol{\mathcal{O}}^-$, with the two sets being mutually exclusive. The individual-level net benefit thereby turns into a function of $\boldsymbol{o}$, $\boldsymbol{\mathcal{O}}^-$, and $\boldsymbol{\theta}$. In state $D$, patients are given their optimal interventions based on $\boldsymbol{o}$ with perfect parameter information. The value of state $D$ is 
\begin{equation} \label{eq:D}
    D = \mathbb{E}_{\boldsymbol{\theta}} \Bigg[ \mathbb{E}_{\mathcal{\boldsymbol{o}} \mid \boldsymbol{\theta}} \Big [\max_d\ \mathbb{E}_{\boldsymbol{\mathcal{O}}^- \mid \mathcal{\boldsymbol{o}}, \boldsymbol{\theta}} \left[nb_d(\mathcal{\boldsymbol{o}}, \boldsymbol{\mathcal{O}}^-, \boldsymbol{\theta})\right] \Big] \Bigg]
\end{equation} 
In state $C$, individuals receive optimal interventions based on $\boldsymbol{o}$, but parameter uncertainty remains. The value of state $C$ is 
\begin{equation} \label{eq:C}
    C = \mathbb{E}_{\mathcal{\boldsymbol{o}}} \Bigg[ \max_d\ \mathbb{E}_{\boldsymbol{\theta} \mid \mathcal{\boldsymbol{o}}} \Big[ \mathbb{E}_{\boldsymbol{\mathcal{O}}^- \mid \boldsymbol{o}, \boldsymbol{\theta}} \big[nb_d(\mathcal{\boldsymbol{o}}, \boldsymbol{\mathcal{O}}^-, \boldsymbol{\theta})\big] \Big] \Bigg].
\end{equation}
The values of states C and D depend on the perfectly predicted outcome components $\boldsymbol{o}$. The choice of $\boldsymbol{o}$ varies with research questions under consideration. For example, $\boldsymbol{o}$ could be the occurrence of a side effect (binary) or disease-free survival time before relapse (continuous). If $\boldsymbol{o}$ contains continuous variables, the expected net benefit conditional on $\boldsymbol{o}$, $\mathbb{E}_{\boldsymbol{\mathcal{O}}^- \mid \mathcal{\boldsymbol{o}}, \boldsymbol{\theta}}[\cdot]$, can be estimated using regression, including parametric and nonparametric models. \cite{strong2014estimating, Heath2022} If $\boldsymbol{o}$ contains discrete variables, the conditional expectation $\mathbb{E}_{\boldsymbol{\mathcal{O}}^- \mid \mathcal{\boldsymbol{o}}, \boldsymbol{\theta}}[\cdot]$ can be computed as the means of the simulated individual-level net benefits for each level of the outcome. \cite{Heath2022} Because states $C$ and $D$ assume perfect prediction for selected outcome components, their values are bounded above by the corresponding values for states $E$ and $F$ (c.f. supplement). 

The difference between states $D$ and $A$ gives the maximum value of research exploring currently unknown characteristics related to the heterogeneity in specific outcome components, which simultaneously resolves the unexplained heterogeneity in $\boldsymbol{o}$ and parameter uncertainty. We define this quantity as the Expected Value of Perfect Outcome Prediction (EVPOP). $C - A$ is the Value of Perfect Outcome Prediction under current parameter uncertainty (VPOP$_{cur}$). Comparing VPOP$_{cur}$ across choices of $\boldsymbol{o}$ identifies the outcome components whose underlying unexplained heterogeneity offers the greatest value for future research with current parameter information. When VPOP$_{cur}$ is substantially higher than EVPI (i.e., $B - A$), future research resolving unexplained heterogeneity in $\boldsymbol{o}$ should be prioritized over parameter uncertainty. We further define $D - C$ as the Outcome Conditional-EVPI (EVPI$_{\boldsymbol{o}}$), which quantifies the value of further research that reduces parameter uncertainty, assuming patients receive the optimal intervention based on perfectly predicted $\boldsymbol{o}$.

\subsection*{Extending the Counterfactual Dependence Structure}
To compute the values in the EVoUH framework, we must determine the optimal net benefit for each individual, which requires an assumption about the counterfactual. As such, specifying the counterfactual is a fundamental part of calculating research value using the EVoUH framework. In the VoUH framework, which only considers two interventions, counterfactual outcomes were defined using a correlation vector $\boldsymbol{\rho}$ by varying the assumed correlation between the two interventions. \cite{Heath2022} However, when more than two interventions are included, a scalar correlation, $\boldsymbol{\rho}$, is insufficient because the entire correlation structure between all interventions is required. Thus, an alternative approach to defining the counterfactual is required.

Even among individuals with homogeneous observed characteristics, health economic outcomes of interventions may remain dependent. We represent this dependence across interventions using a shared latent factor $\boldsymbol{U}$. The latent factor does not represent a specific unmeasured patient characteristic. Instead, it serves as a computational device for inducing dependence among intervention-specific net benefits beyond directly observable patient characteristics while preserving their marginal distributions. \cite{ruschendorf2009distributional, krupskii2013factor} For computational convenience, we assume that $\boldsymbol{U}$ follows a normal distribution. Rather than directly specifying a full correlation matrix across interventions, we define the dependence structure across intervention-specific net benefits through intervention-specific correlations with the latent factor $\boldsymbol{U}$, \cite{ruschendorf2009distributional, krupskii2013factor} denoted by $\boldsymbol{\rho}_{d_k,\boldsymbol{U}}$. Here, $d_{k}$ denote the $k$th intervention included in research, $d_k \in \{d_1, \dots, d_D \}$.  We sample intervention-specific correlations with $\boldsymbol{U}$ from $[-1,1]$ to explore alternative dependence structures. Specifically, we reorder intervention-specific net benefits based on their sampled correlations with $\boldsymbol{U}$, then estimate the pairwise correlations between interventions, $\boldsymbol{\rho}_{d_{k},d_{l}}$, from the re-ordered data. The values of $\boldsymbol{\rho}_{d_k,\boldsymbol{U}}$ are fixed for a given scenario and the individual-level net monetary benefits at each value of $\boldsymbol{\theta}$ are reordered using the ’$\texttt{SimJoint}$’ package in R (version 4.3.1) to compute the measures in our framework. \cite{SimJoint, Heath2022} 

\subsection*{Computational Efficiency} 
As counterfactuals are fundamental to the calculation of most research values in the EVoUH framework (i.e., EMVoH, MVoH$_{cur}$, EVPOP, VPOP$_{cur}$, and EVPI$_{\boldsymbol{o}}$), the computational efficiency of obtaining these counterfactuals through re-ordering is an important consideration for implementing the framework. Unfortunately, the reordering step is computationally intensive (Table S3). The computational efficiency of this procedure depends on both the number of iterations in the PA process $n_{\text{PA}}$ and the number of individuals included in the DAMs $n_i$: \cite{liu2019simjoint}
    \begin{equation} \label{eq:computational_efficiency}
    O(n_{\text{PA}} \times m \cdot n_i \cdot log(n_i)), 
    \end{equation} 
where $m$ is the number of internal iterations in the ’$\texttt{SimJoint}$’ determined by the setting of ’$\texttt{convergenceTail}$’. \cite{SimJoint} As expected, computational complexity and runtime show a positive correlation (Figure S1). When computational resources are limited, one may need to trade off the stability of the estimated results against the runtime of the functions in the \texttt{Simjoint} package. 

Monte Carlo standard errors (MCSEs) of the values of the states in the EMVoH framework can be calculated using the pigeonhole bootstrap. \cite{Owen2007} Assuming that individuals included in the DAMs and the parameter sets used in PA are independent, MCSEs have the following relationship with $n_i$, and $n_{\text{PA}}$ is (c.f. supplement):  
\begin{align}
    \text{SE}(A), \text{SE}(B), \text{SE}(E), \text{SE}(F) & \leq \frac{1}{ n_i \cdot n_{\text{PA}} } \sqrt{ \sum_{i = 1}^{n_i} \sum_{j = 1}^{n_{\text{PA}}}  \Big[\sum_{d \in \boldsymbol{\mathcal{D}}} \text{Var} \big( nb_{ij}(d) \big) \Big] } \label{eq:se} 
\end{align} 
where $nb_{ij}(d)$ is the net monetary benefit for individual $i$ under intervention $d$ estimated using the PA parameter set $j$. The upper bounds of MCSEs follow relative fixed-width stopping rules for Markov chain Monte Carlo (MCMC) \cite{gong2016practical}: 
    \begin{equation}
        \frac{\text{MCSE}(\hat{\gamma})}{|\hat{\gamma}|} < \epsilon, \quad \epsilon \in \{0.01, 0.05\}, \label{eq:stopping_rule}
    \end{equation} 
where $\hat{\gamma}$ denotes the simulated values for the states within the EVoUH framework. 

Together, given limited computational resources, the upper bound for MCSEs of the states should be estimated first (Eq.~\ref{eq:se} and ~\ref{eq:stopping_rule}). We then suggest choosing a sufficiently large sample size $n_i$ to ensure stable MCSE, as $n_i$ drives the computational time (Eq.~\ref{eq:computational_efficiency}). Based on $n_i$, we recommend a smaller $n_{\text{PA}}$ that yields the lowest computational time when the computational resources are limited. 

\subsection*{Illustrative Example: Sick-Sicker Microsimulation Model}
We now use an example to demonstrate the proposed framework and methods in research prioritization. The previously developed Sick-Sicker model (Figure~\ref{fig:msm}) is a state-transition microsimulation model with four health states, Healthy (H), Sick (S$_1$), Sicker (S$_2$), and Death (D). \cite{Krijkamp2018} Arrows between any two health states indicate that individuals can transition between the states at each cycle. 
\begin{center}
    \includegraphics[width=0.8\textwidth]{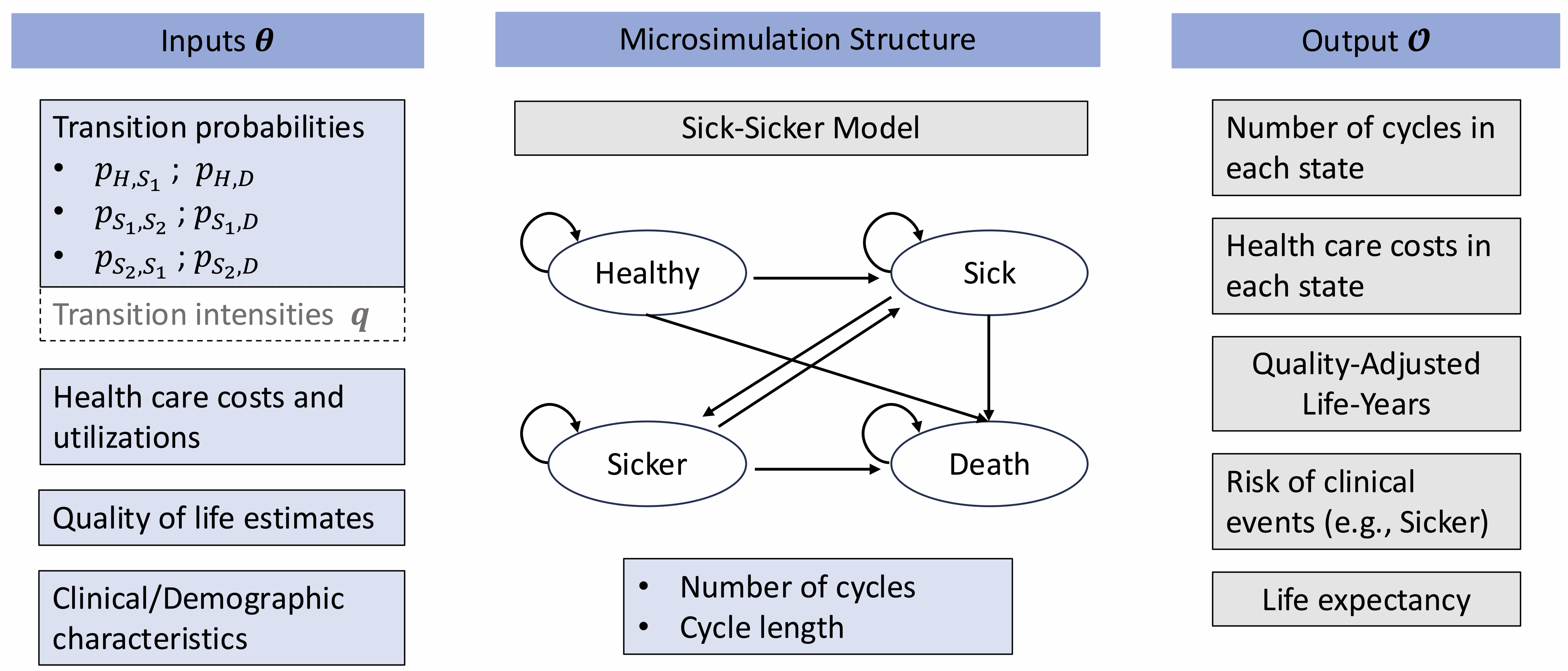} 
    \captionof{figure}{Four-state sick-sicker microsimulation model.\cite{Krijkamp2018}}
    \label{fig:msm}
\end{center} 
We assume that the Sick-Sicker model simulates the progression of a hypothetical disease in patients with homogeneous observed characteristics. We compare three interventions, standard of care (SoC), treatment 1 (Trt1), and treatment 2 (Trt2) (i.e., $d \in \{ \text{SoC, Trt1, Trt2} \}$), and use a cycle length of 1 year and a model horizon of 60 cycles to simulate lifetime trajectories. We denote the model parameters (e.g., between health state transition probabilities/intensities, health state-specific costs, and effectiveness) by $\boldsymbol{\theta}$, which follow different probability distributions across interventions. Table S1 and S2 summarize the specific parameter distributions of $\boldsymbol{\theta}$. 

Individual-level net monetary benefits under intervention $d$ are calculated using the following function. 
\begin{align} 
     nb_d &= \lambda \left[ \sum_{t=1}^{n_t} \Big(I\{ S_t = H \} \cdot e_H + I\{ S_t = S_1 \} \cdot e_{s_1} + I\{ S_t = S_2 \} \cdot e_{s_2}\Big) \times D_{e, t} \omega_t \right] -  \label{eq:my_equation} \\
     &\left[ \sum_{t=1}^{n_t} \Big(I\{ S_t = H \} \cdot c_H + I\{ S_t = S_1 \} \cdot c_{s_1} + I\{ S_t = S_2 \} \cdot c_{s_2} \Big) \times D_{c, t} \omega_t + I\{ S_t = D \} \cdot c_D \right], \notag 
\end{align} 
where $e_{\text{state}}$ and $c_{\text{state}}$ are the effectiveness and costs for a specific health state, and $I\{ S_t = \text{state} \}$ indicates the health state for each individual at cycle $t$. $D_{c, t}$ and $D_{e, t}$ denote the discount factors for costs and effectiveness. $\omega_t$ is the within-cycle correction weight. Thus, the outcome components $\boldsymbol{\mathcal{O}}$ in the Sick-Sicker model comprise indicators of the health states entered by the patient, corresponding health state-specific costs, and health effects at each cycle of the patient's disease trajectory. $\boldsymbol{\mathcal{O}}$ can be decomposed in different ways to obtain different components $\boldsymbol{o}$. For example, we could record whether individuals enter each state or the number of cycles individuals stay in each state. 

We included 10,000 individuals in the microsimulation model and used 10,000 iterations in the PA to capture individual heterogeneity and parameter uncertainty. To account for the counterfactual results, 10,000 values for the shared latent dependence factor $\boldsymbol{U}$ were simulated from $\boldsymbol{U} \sim N(0, 1)$. We randomly simulated 20 correlations between $\boldsymbol{U}$ and the net benefits of each intervention.

\section*{Results}
\subsection*{General Research Direction: From Population to Individualized Level}
We first estimated the EMVoH to assess the value of conducting further research (Eq.~\ref{eq:emvoh}). Fig.~\ref{fig:counterfactual_abef} shows that the EMVoHs remain substantial across between-intervention correlations, ranging from $\$$60,646 to $\$$94,314. EMVoHs gradually increase as correlations decrease, suggesting a higher value of further research when health economic outcomes differ more substantially across interventions. Together, they suggest that future research is important in order to improve decision making compared to the current optimal decision. 

To identify valuable research directions, we then calculated and compared the EVPI and MVoH$_{cur}$ based on Eqs.~\ref{eq:mvoh_cur} and~\ref{eq:evpi} using the simulated individual-level net benefit (c.f., supplement). 
\begin{center}
    \includegraphics[width=\textwidth]{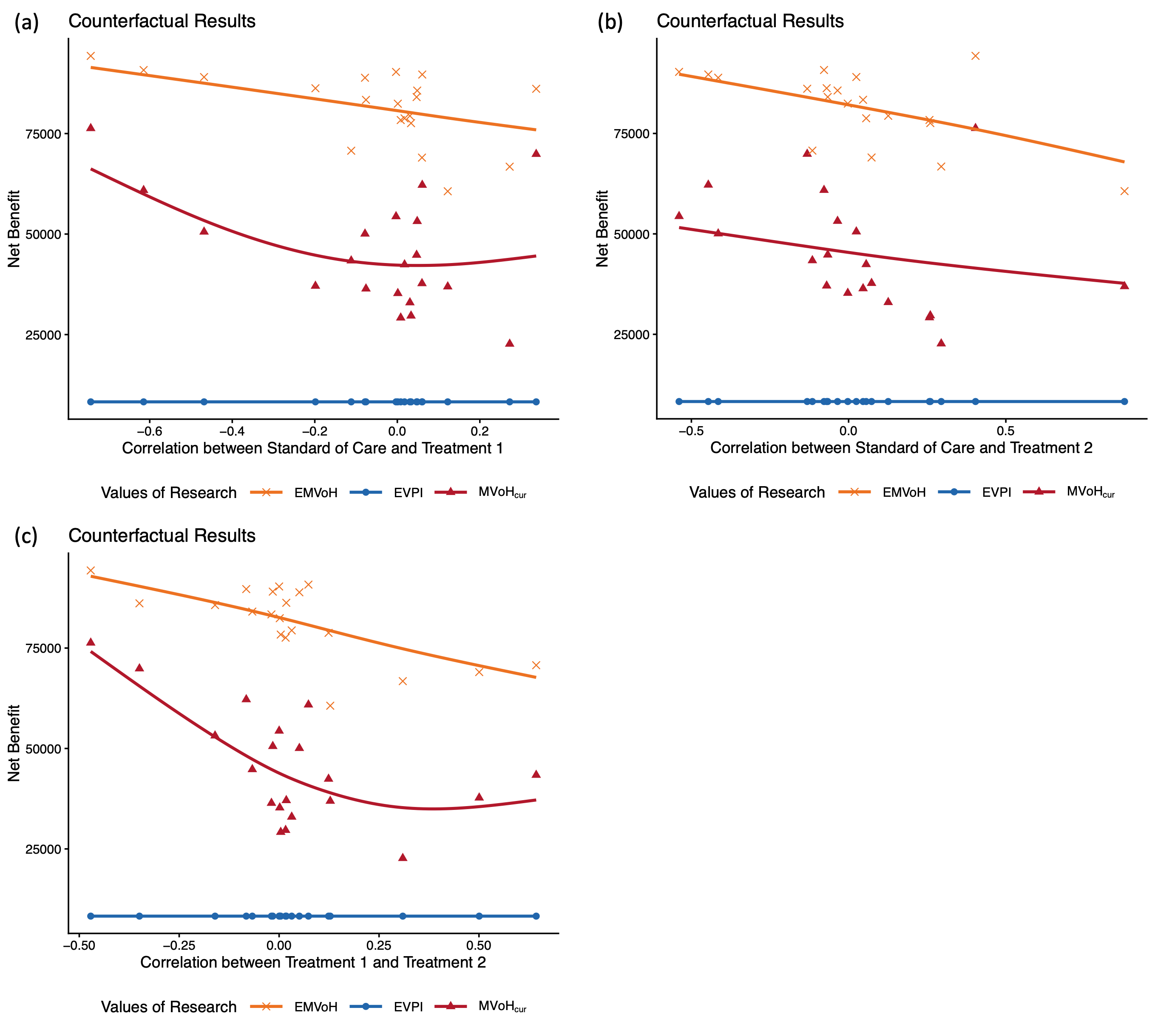} 
    \captionof{figure}{The EMVoH, EVPI, and MVoH under current parameter intervention across the correlations between the individual-level net benefit of (a) standard of care and treatment 1, (b) standard of care and treatment 2, and (c) treatment 1 and treatment 2.}
\label{fig:counterfactual_abef}
\end{center} 
MVoH$_{cur}$ shows a similar trend to the EMVoH across correlations, ranging from $\$$22,676 to $\$$76,297 (Figure~\ref{fig:counterfactual_abef}). In contrast, EVPI remains consistent at $\$$8,271, substantially lower than the MVoH$_{cur}$ at all correlations. In this case, when resources are limited, research exploring novel precision health strategies (e.g., identifying new biomarkers predictive of disease onset) should be prioritized over research reducing parameter uncertainty (e.g., getting a more precise population-average treatment effect). 

Following the general research direction of addressing unexplained heterogeneity, we further explore the values of studies targeting different outcome components.

\subsection*{Prioritizing research for specific outcome component}
To identify the outcome component whose related unexplained heterogeneity drives the effect, we selected four different $\boldsymbol{o}$: duration of life ($\boldsymbol{n_l}$), time in Healthy state ($\boldsymbol{n}_{\text{H}}$), entry into Sicker state ($\boldsymbol{I}\{ \text{state} = \text{S}_2 \}$), and total costs ($\boldsymbol{C}$). The corresponding $\boldsymbol{\mathcal{O}}^-$ are the remaining outcome components in $\boldsymbol{\mathcal{O}}$. Figure~\ref{fig:counterfactual_abcd} shows that EVPOPs remain substantial across correlations for all potential $\boldsymbol{o}$, indicating that further investigations of all four $\boldsymbol{o}$ could meaningfully reduce decision uncertainty.  
\begin{center}
    \includegraphics[width=\textwidth]{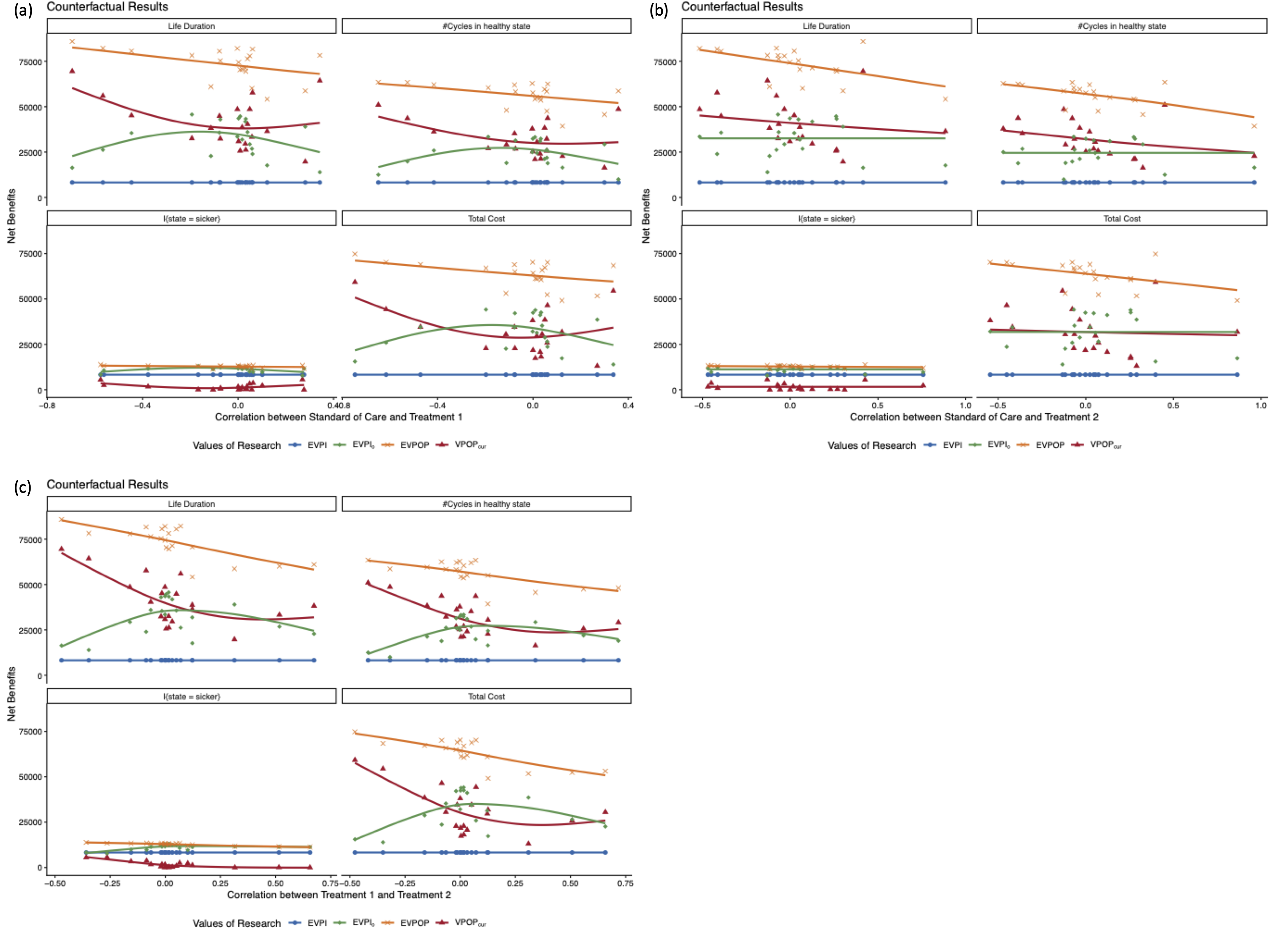} 
    \captionof{figure}{The EVPOP, EVPI, and VPOP under current parameter information, and outcome conditional-EVPI against the correlations between the net benefit across (a) standard of care and treatment 1, (b) standard of care and treatment 2, and (c) treatment 1 and treatment 2.}
    \label{fig:counterfactual_abcd}
\end{center} 

VPOP$_{cur}$ values were similar when life durations, time in Healthy state and total costs could be perfectly predicted, ranging from $\$$19,719 to $\$$69,514, $\$$16,394 to $\$$50,959, and $\$$13,072 to $\$$59,258 across all correlations, respectively. These values were substantially higher than the EVPI ($\$$8,271). Therefore, research resolving the unexplained heterogeneity underlying these outcome components may be prioritized under limited research budgets. In contrast, when the research interest $\boldsymbol{o}$ is entry into Sicker state, the corresponding VPOP$_{cur}$ remained lower than the EVPI across all counterfactual scenarios, ranging from $\$$0 to $\$$5,634 (Figure~\ref{fig:counterfactual_abcd}). In other words, research reducing parameter uncertainty will likely be more valuable than exploring new biomarkers for predicting $\boldsymbol{I}\{ \text{state} = \text{S}_2 \}$. In this case, EVPPI should be estimated to identify valuable research directions. \cite{jackson2022value}

We also calculated the EVPI$_{\boldsymbol{o}}$, which remained high across correlations for all outcomes considered. Therefore, reducing parameter uncertainty may be considered once the unexplained heterogeneity is resolved.

\subsection*{Computational Efficiency for Counterfactual Estimation}
Although we had sufficient computational resources to obtain the results in this example with 10,000 individuals and PA iterations, we still used the simulation results to assess how to select $(n_i, n_{\text{PA}})$ under limited computational resources. We measured the computational time to obtain counterfactual outcomes using the \texttt{SimJoint} package by running simulations with varying numbers of individuals (1,000 to 10,000) and PA iterations (1,000 to 10,000). As the number of included individuals and PA iterations increased, computational time increased, while the MCSEs of $A$, $B$, $E$ and $F$ decreased. 

The MCSEs of the expected values became insensitive to the number of individuals included in the microsimulation model once it reached 6,000  (Figure S2). The corresponding MCSEs were below 0.01 of the estimated values for each state and EMVoH, thereby satisfying the relative fixed-width stopping rule (Eq.~\ref{eq:stopping_rule}). When the MCSE became insensitive at 6,000 individuals, computational time ranged from 200 to 800 seconds (Figure S3). Thus, $n_i=6,000$ and $n_{\text{PA}} = 5,000$ could be selected to maintain a similar MCSE to the results in this study while reducing computational burden. 

\section*{Discussion}
We developed a framework to guide research prioritization in an era of precision health, when diverse research directions and objectives are possible while research budgets and resources remain limited. To achieve this, we developed the Expected Value of Unexplained Heterogeneity framework, which encompasses both concepts of parameter uncertainty and unexplained heterogeneity. Differences in values of each ``state of information'' in the framework estimate the value of research targeting different sources of uncertainty. Generalizing previous work on the Value of Unexplained Heterogeneity, \cite{Heath2022} we used a latent dependence factor to allow counterfactual estimation when more than two interventions are included in research. To reduce the computational burden of counterfactual estimation, we suggested using the relative fixed-stopping rule with MCSE boundaries to guide the choice of PA iterations and individuals in decision-analytic models. This framework allows direct comparisons across research directions, ranging from reducing parameter uncertainty (e.g., more precise estimation of a new drug’s population-average treatment effect) to exploring unexplained heterogeneity (e.g., identifying new biomarkers for predicting cancer relapse). 

Several limitations of the proposed framework should be addressed in future work. First, we assume that individual-level outcomes can be perfectly predicted. Although the framework identifies which research direction and outcome component to prioritize, individual-level outcomes are rarely perfectly predictable. A natural follow-up question is whether a specific study design for particular outcome components may be beneficial. The Value of Subgroup (VoS), established in prior work, quantifies the extent to which decision uncertainty is driven by uncertainty in a study design's latent subgroup composition, assuming a deterministic parameter vector for each scenario. \cite{Heath2022} Taking the VoS as a reference, future research should extend the EVoUH framework to settings that estimate and compare the values of specific studies. 

Secondly, we constrained our settings to individuals having homogeneous observed characteristics. However, in real-world studies, patients generally have different observed characteristics. Thus, future studies should extend the EVoUH framework to settings which also consider explainable heterogeneity. Methods in the Value of Heterogeneity framework may be adapted for this purpose. \cite{Espinoza2014} Ideally, this extension would guide research prioritization in real-world settings, where diverse research directions exist.

Thirdly, although we provided a solution for limited computational resources, further research could improve the computational efficiency of estimating counterfactual results by using a novel method that accounts for the correlation matrix rather than using the '\texttt{SimJoint}' package. Finally, as the framework was demonstrated using a hypothetical example, further application to real-world decision problems is needed to confirm its practical applicability.

In conclusion,  our framework enables direct, quantitative comparisons of value across different research directions, identifying studies with the greatest potential gains. It serves as a practical tool for selecting the highest-value studies and estimating potential gains to support personalized research proposals.

\clearpage
\setcounter{table}{0}
\setcounter{figure}{0}
\renewcommand{\thetable}{S\arabic{table}}
\renewcommand{\thefigure}{S\arabic{figure}}

\begin{center}
    \textbf{\large Supplementary Material}
\end{center}

\section*{Table}
{\begin{table}[H]
    \centering 
    \vspace{-0.7\baselineskip} 
    \renewcommand{\arraystretch}{1.0} 
    
    \begin{singlespace} 
        \caption{The population-level parameter values for our individual-level decision model, microsimulation model, comparing the three interventions (standard of care, treatment 1, and treatment 2) of interest.} 
        \label{tab:fixed_parameter}
    \end{singlespace}  
    
    \begin{threeparttable}
        \begin{tabular}{l >{\raggedright\arraybackslash}p{11.5cm} l}
            \toprule 
            Parameter & Description & Value \\ 
            \midrule 
            $n_l$ & Number of cycles & 60 \\ 
            $l_c$ & Length of each cycle & 1 \\ 
            $n_i$ & Number of individuals included in the microsimulation models & 10,000 \\ 
            $c_{\text{SoC}}$ & Cost of standard of care during each cycle ($\$$) & 10 \\ 
            $c_{\text{trt}_1}$ & Cost of Treatment 1 during each cycle ($\$$) & 800 \\ 
            $c_{\text{trt}_2}$ & Cost of Treatment 2 during each cycle ($\$$) & 1100 \\ 
            $D_{c, t}$ & Discounting factor for costs & 0.03 \\ 
            $D_{e, t}$ & Discounting factor for utilities (Quality-Adjusted Life Year) & 0.03 \\ 
            \bottomrule 
        \end{tabular}
    %\begin{tablenotes} 
        %\small 
        %\item Note: 
    %\end{tablenotes} 
    \end{threeparttable}
\end{table}
}

\clearpage
\begin{center}
\vspace{-0.7\baselineskip}
\renewcommand{\arraystretch}{1.0}
\begin{ThreePartTable} 
    \begin{longtable}{l >{\raggedright\arraybackslash}p{10cm} l} 
        \caption{Population-level Parameter Distributions} 
        \label{tab:parameter_distribution}\\ 
        \toprule 
        Parameter & Description & Distribution \\ 
        \midrule 
        $p_{\text{H}, \text{S}_1}$ & Transition probability from healthy state to sick state & $Beta(\alpha_{\text{H}, \text{S}_1}, \beta_{\text{H}, \text{S}_1})$ \\ 
        $p_{\text{H}, \text{D}}$ & Transition probability from healthy state to death & $Beta(\alpha_{\text{H}, \text{D}}, \beta_{\text{H}, \text{D}})$  \\ 
        $q_{\text{S}_1, \text{S}_2}^{\text{SoC}}$ & Transition intensity from sick state to sicker state with standard of care & $\Gamma(\alpha_{\text{S}_1, \text{S}_2}^{\text{SoC}}, \lambda_{\text{S}_!, \text{S}_2}^{\text{SoC}})$ \\ 
        $q_{\text{S}_1, \text{D}}^{\text{SoC}}$ & Transition intensity from sick state to death with standard of care & $\Gamma(\alpha_{\text{S}_1, \text{D}}^{\text{SoC}}, \lambda_{\text{S}_!, \text{D}}^{\text{SoC}})$ \\ 
        $p_{\text{S}_2, \text{S}_1}^{\text{SoC}}$ & Transition probability from sicker state back to sick state with standard of care & $Beta(\alpha_{\text{S}_2, \text{S}_1}^{\text{SoC}}, \beta_{\text{S}_2, \text{S}_1}^{\text{Soc}})$ \\ 
        $p_{\text{S}_2, \text{D}}^{\text{SoC}}$ & Transition probability from sicker state back to sick state with standard of care & $Beta(\alpha_{\text{S}_2, \text{D}}^{\text{SoC}}, \beta_{\text{S}_2, \text{D}}^{\text{SoC}})$  \\ 
        $q_{\text{S}_1, \text{D}}^{\text{Trt1}}$ & Transition intensity from sick state to death with treatment 1 & $\Gamma(\alpha_{\text{S}_1, \text{D}}^{\text{Trt1}}, \lambda_{\text{S}_1, \text{D}}^{\text{Trt1}})$ \\ 
        $q_{\text{S}_2, \text{S}_1}^{\text{Trt1}}$ & Transition intensity from sicker state back to sick state with treatment 1 & $\Gamma(\alpha_{\text{S}_2, \text{S}_1}^{\text{Trt1}}, \lambda_{\text{S}_2, \text{S}_1}^{\text{Trt1}})$ \\ 
        $p_{\text{S}_2, \text{S}_1}^{\text{Trt1}}$ & Transition probability from sicker state back to sick state with treatment 1 & $Beta(\alpha_{\text{S}_2, \text{S}_1}^{\text{Trt1}}, \beta_{\text{S}_2, \text{S}_1}^{\text{Trt1}})$ \\ 
        $p_{\text{S}_2, \text{D}}^{\text{Trt1}}$ & Transition probability from sicker state back to sick state with treatment 1 & $Beta(\alpha_{\text{S}_2, \text{D}}^{\text{Trt1}}, \beta_{\text{S}_2, \text{D}}^{\text{Trt1}})$  \\ 
        $q_{\text{S}_1, \text{D}}^{\text{Trt2}}$ & Transition intensity from sick state to death with treatment 2 & $\Gamma(\alpha_{\text{S}_1, \text{D}}^{\text{Trt2}}, \lambda_{\text{S}_1, \text{D}}^{\text{Trt2}})$ \\ 
        $q_{\text{S}_2, \text{S}_1}^{\text{Trt2}}$ & Transition intensity from sicker state back to sick state with treatment 2 & $\Gamma(\alpha_{\text{S}_2, \text{S}_1}^{\text{Trt2}}, \lambda_{\text{S}_2, \text{S}_1}^{\text{Trt2}})$ \\
        $p_{\text{S}_2, \text{S}_1}^{\text{Trt2}}$ & Transition probability from sicker state back to sick state with treatment 2 & $Beta(\alpha_{\text{S}_2, \text{S}_1}^{\text{Trt2}}, \beta_{\text{S}_2, \text{S}_1}^{\text{Trt2}})$ \\ 
        $p_{\text{S}_2, \text{D}}^{\text{Trt2}}$ & Transition probability from sicker state back to sick state with treatment 2 & $Beta(\alpha_{\text{S}_2, \text{D}}^{\text{Trt2}}, \beta_{\text{S}_2, \text{D}}^{\text{Trt2}})$  \\ 
        $c_{\text{H}}$ & The cost of health state in each cycle  & Lognrmal$_c(\mu_{\text{H}}, \sigma_{\text{H}}^2)$\\ 
        $c_{\text{S}_1}$ & The cost of sick state in each cycle & Lognrmal$_c(\mu_{\text{S}_1}, \sigma_{\text{S}_1}^2)$\\  
        $c_{\text{S}_2}$ & The cost of sicker state in each cycle & Lognrmal$_c(\mu_{\text{S}_2}, \sigma_{\text{S}_2}^2)$\\ 
        $c_{\text{D}}$ & The one-time cost of death state & Lognrmal$_c(\mu_{\text{D}}, \sigma_{\text{D}}^2)$\\  
        $u_{\text{H}}$ & The utility (Quality-Adjusted Life Year) of health state in each cycle  & Lognrmal$_u(\mu_{\text{H}}, \sigma_{\text{H}}^2)$\\ 
        $u_{\text{S}_1}$ & The utility (Quality-Adjusted Life Year) of sick state in each cycle & Lognrmal$_u(\mu_{\text{S}_1}, \sigma_{\text{S}_1}^2)$\\ 
        $u_{\text{S}_2}$ & The utility (Quality-Adjusted Life Year) of sicker state in each cycle & Lognrmal$_u(\mu_{\text{S}_2}, \sigma_{\text{S}_2}^2)$\\ 
        \bottomrule 
    \end{longtable}
\end{ThreePartTable}
\end{center}

\clearpage
\begin{center}
\vspace{-0.7\baselineskip}
\renewcommand{\arraystretch}{1.0}
    \begin{longtable}{lrrr} 
    \caption{'Rprof()' Results for the Reordering Process in Counterfactual Estimation: sick-sicker model case study}
    \label{tab:rprof_result}\\
     
        \hline 
        Function & Self Time & Self \% & Total Time \\ 
        \hline 
        SimJoint::SJpearson       & 796.16 & 42.71 & 796.20 \\ 
        saveRDS                   & 170.24 & 9.13  & 170.28 \\ 
        merge.data.frame          & 149.98 & 8.05  & 462.84 \\ 
        nzchar                    & 107.98 & 5.79  & 107.98 \\ 
        {[}.data.frame            & 77.48  & 4.16  & 170.42 \\ 
        order                     & 58.62  & 3.15  & 64.62  \\ 
        unlist                    & 56.70  & 3.04  & 56.82  \\ 
        anyDuplicated.default     & 54.58  & 2.93  & 54.58  \\ 
        array                     & 53.20  & 2.85  & 54.10  \\ 
        apply                     & 45.30  & 2.43  & 329.48 \\  
        aperm.default             & 43.86  & 2.35  & 43.88  \\ 
        sort.int                  & 41.96  & 2.25  & 112.10 \\ 
        cbind                     & 40.92  & 2.20  & 168.82 \\ 
        as.vector                 & 40.80  & 2.19  & 40.86  \\ 
        length                    & 18.90  & 1.01  & 18.90  \\ 
        rnorm                     & 10.72  & 0.58  & 11.18  \\ 
        c                         & 10.44  & 0.56  & 10.44  \\ 
        as.data.frame             & 6.78   & 0.36  & 92.92  \\ 
        .rowNamesDF<-             & 6.46   & 0.35  & 7.30   \\ 
        data.frame                & 5.96   & 0.32  & 127.90 \\ 
        \hline 
    \end{longtable}
\end{center}

\newpage
\section*{Figure}
\setcounter{figure}{0}
\renewcommand{\thefigure}{S\arabic{figure}}

\captionsetup{type=figure}
\begin{center}
    \includegraphics[width=0.6\textwidth]{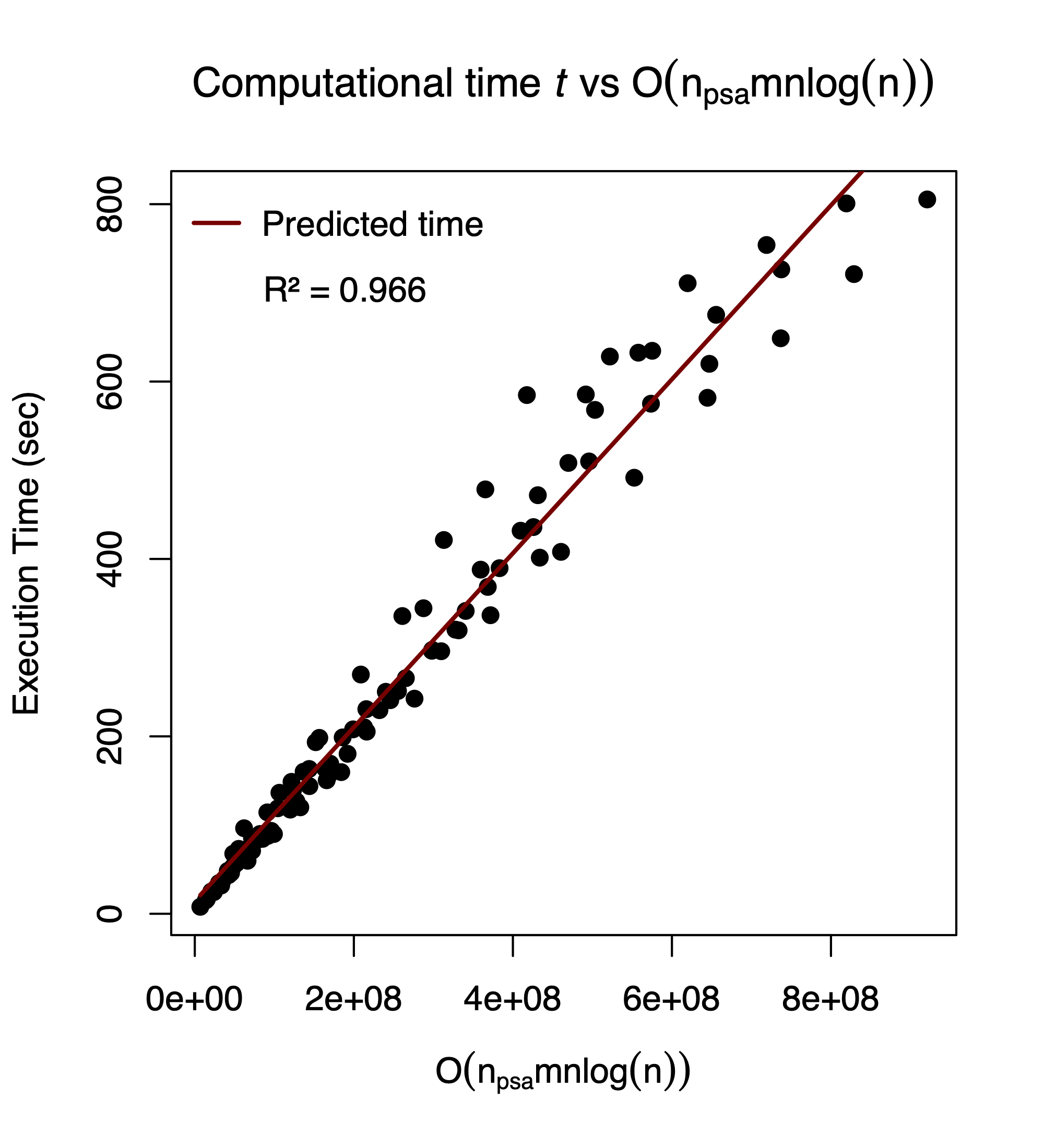} 
    \captionof{figure}{Computational Time versus Computational Complexity} 
    \label{fig:comptime}
\end{center}

\captionsetup{type=figure}
\begin{center}
\begin{longtable}{cc}
\includegraphics[width=0.48\textwidth]{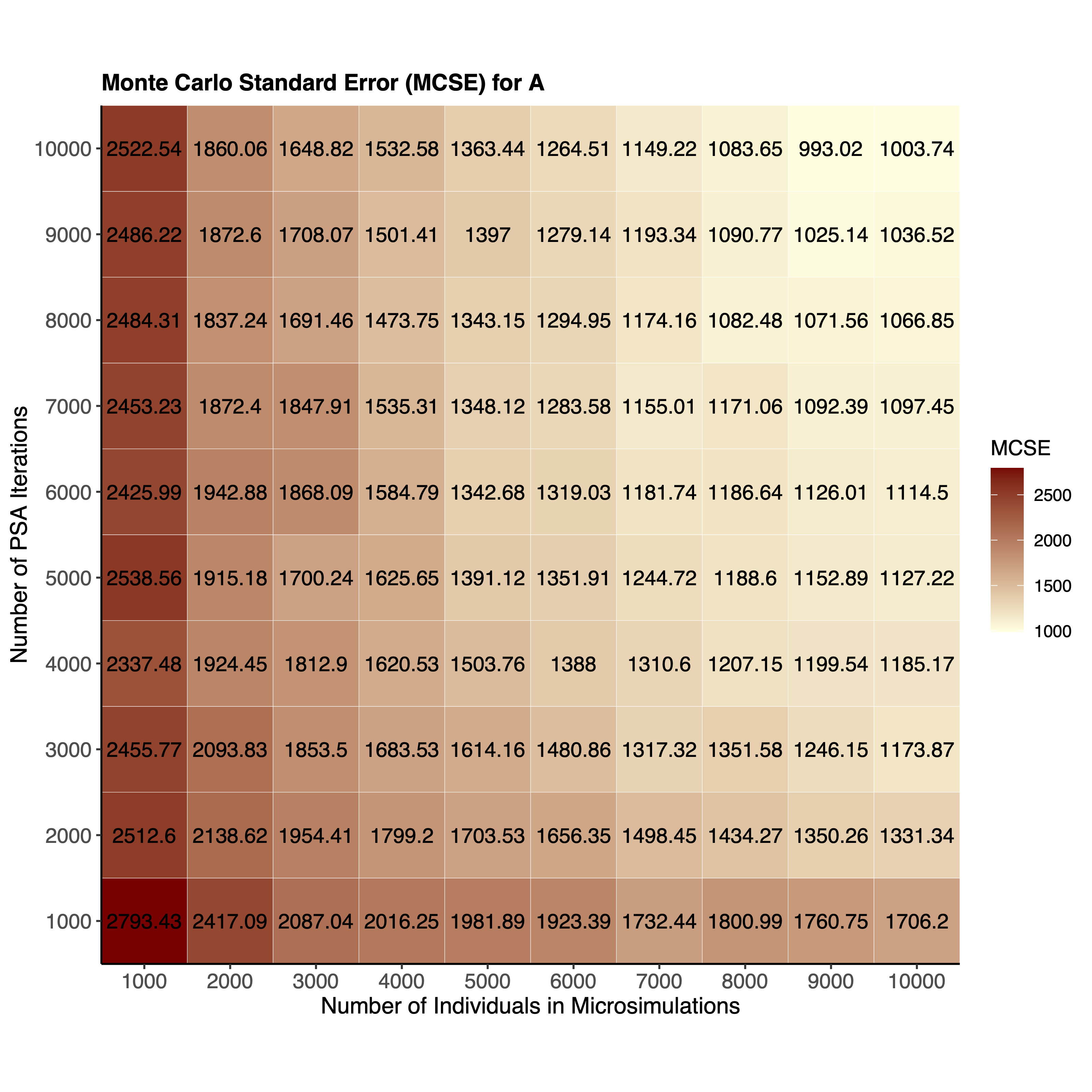} &
\includegraphics[width=0.48\textwidth]{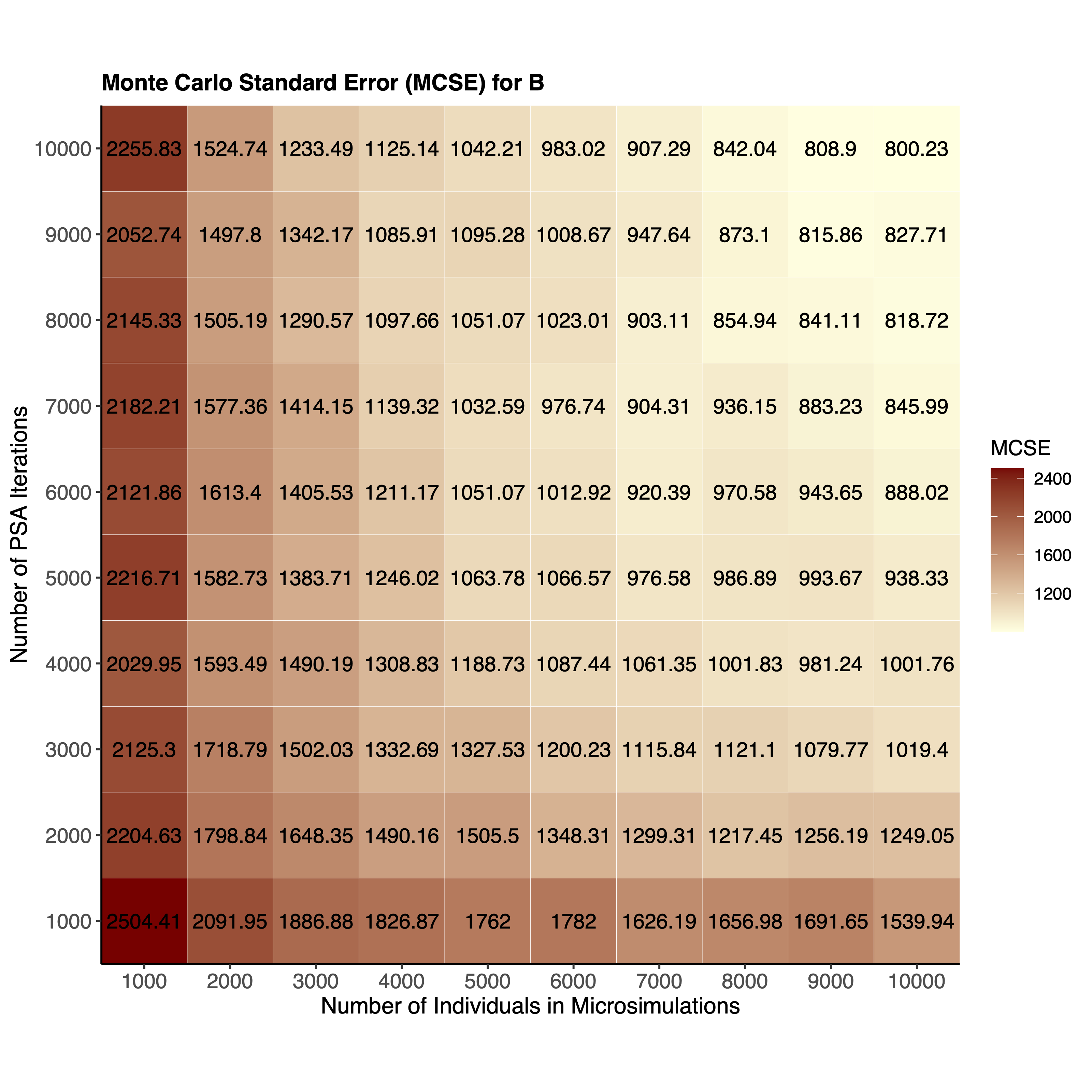} \\[0.5em]
\includegraphics[width=0.48\textwidth]{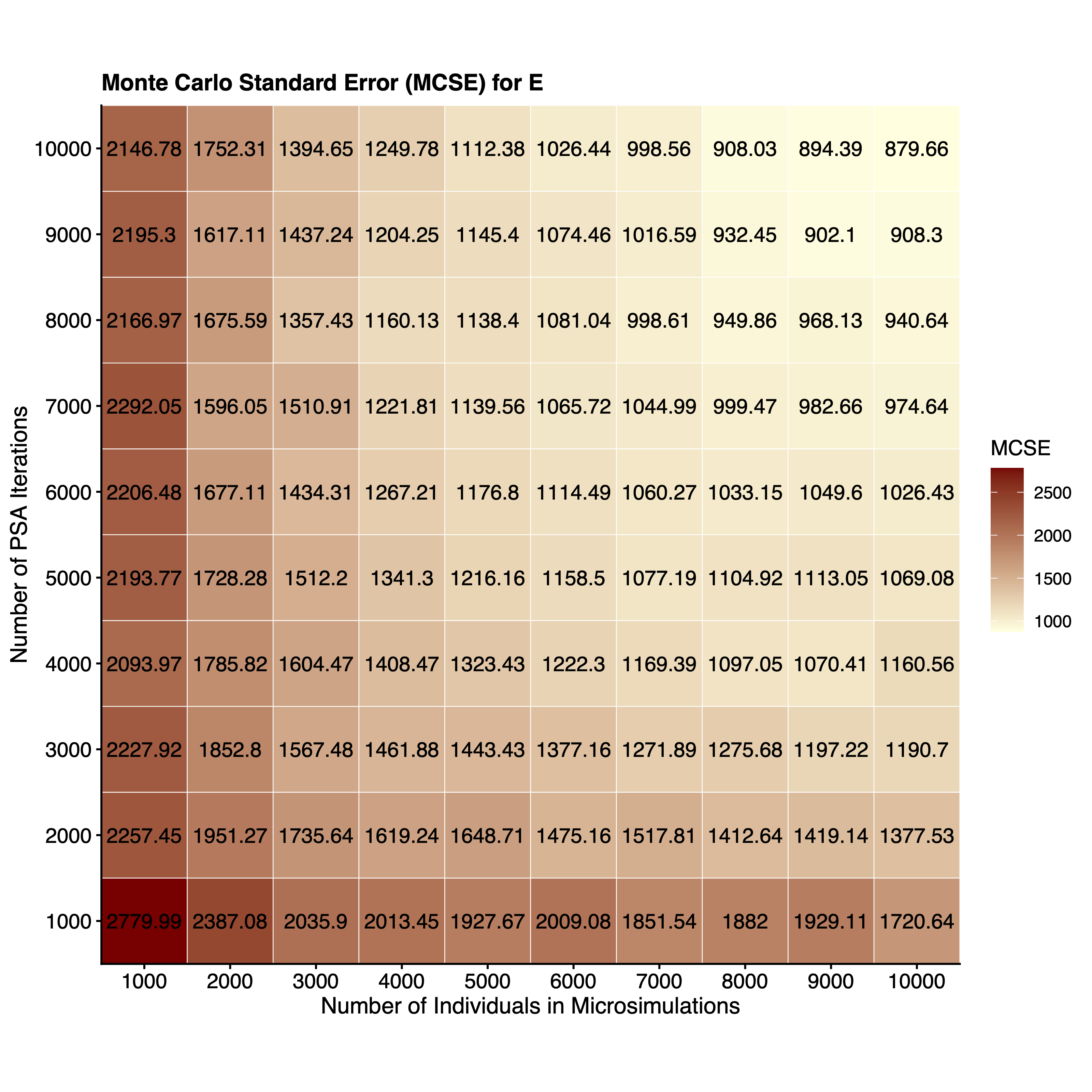} &
\includegraphics[width=0.48\textwidth]{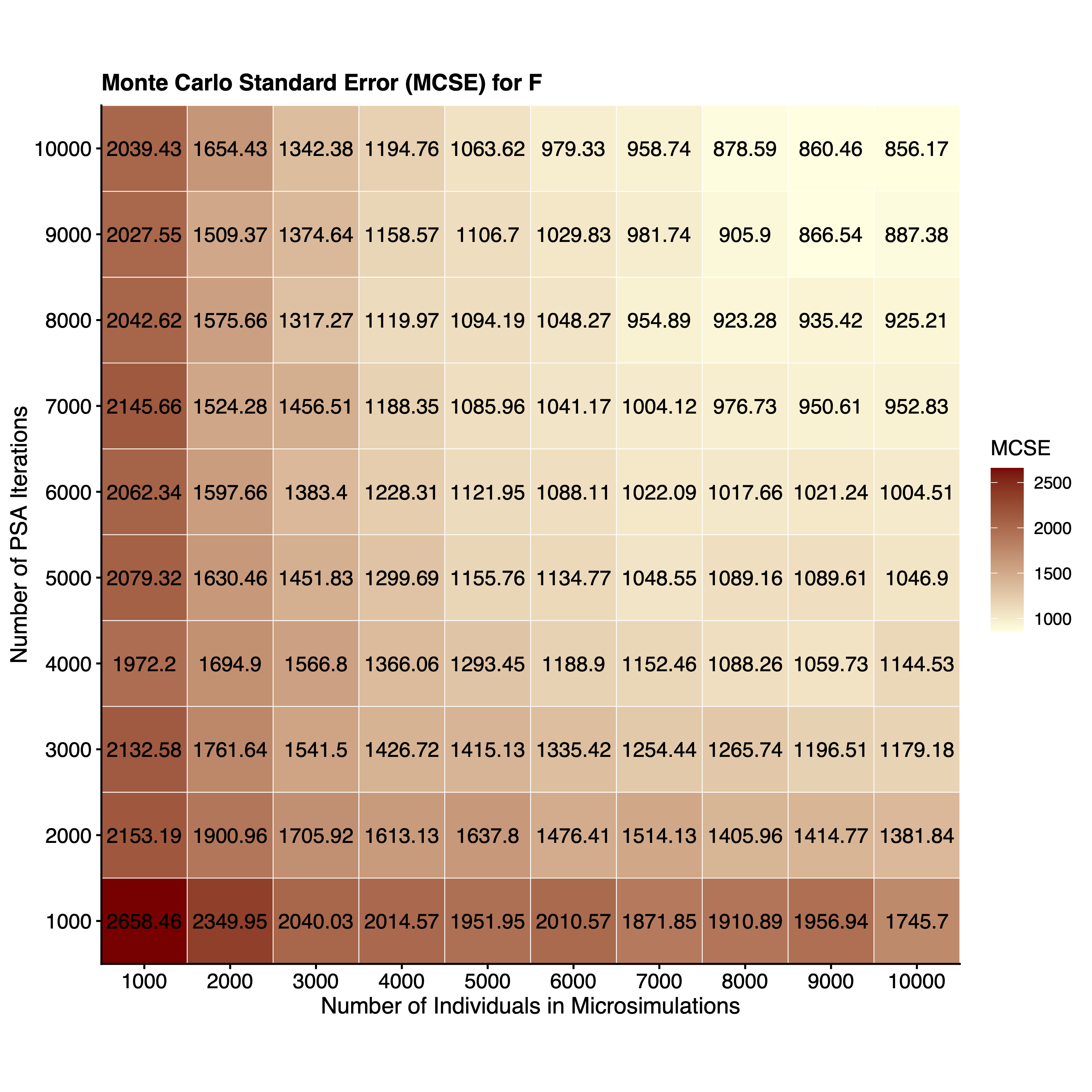} \\[0.5em]
\includegraphics[width=0.48\textwidth]{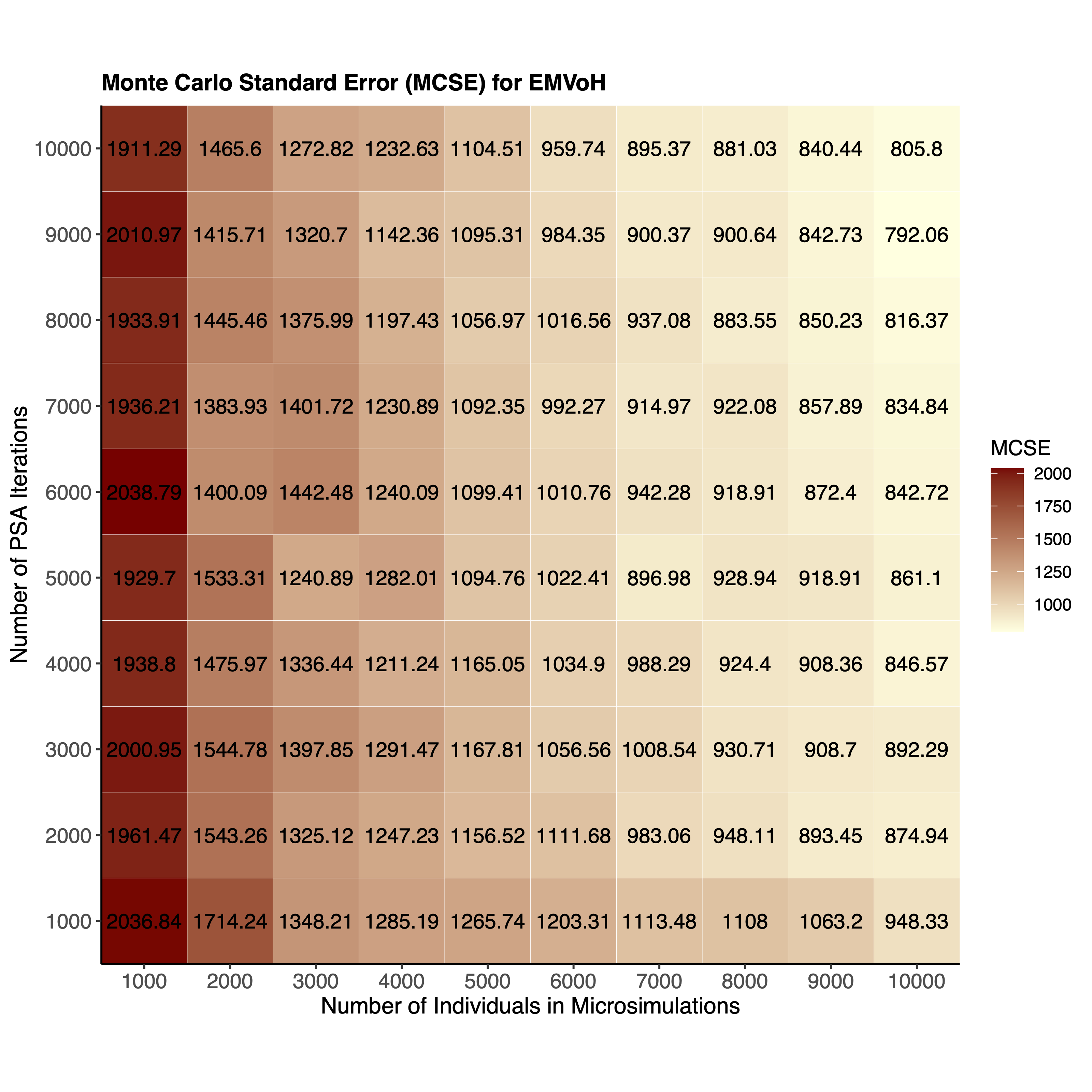} &
\end{longtable}
\end{center}
\vspace{-4em}
\captionof{figure}{Monte Carlo standard errors for the number of simulations included in the microsimulation model and the probabilistic analysis for states $A$, $B$, $E$, $F$, and the expected maximum value of heterogeneity.}
\label{fig:mcse}

\captionsetup{type=figure}
\begin{center}
\begin{longtable}{cc}
\includegraphics[width=0.48\textwidth]{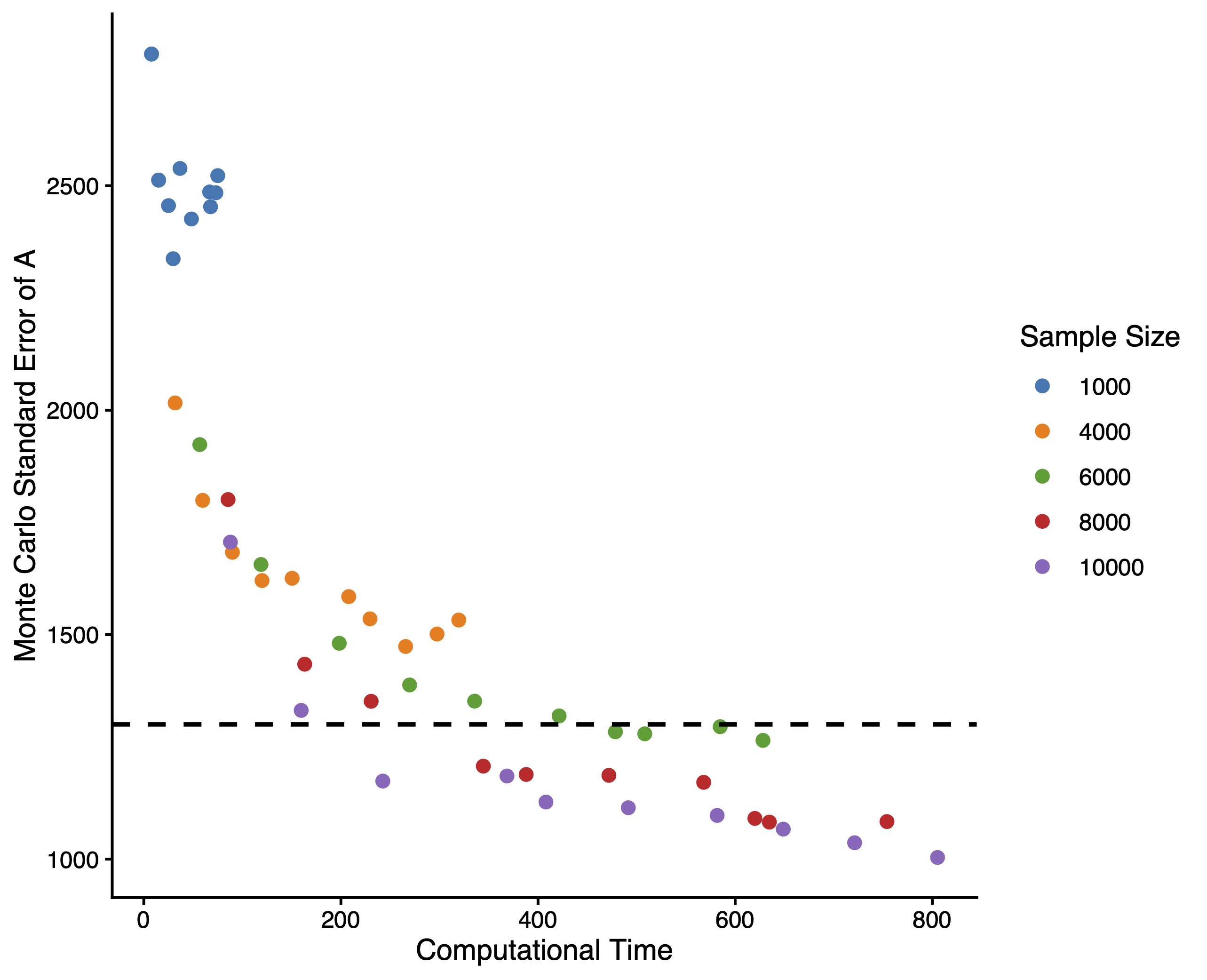} &
\includegraphics[width=0.48\textwidth]{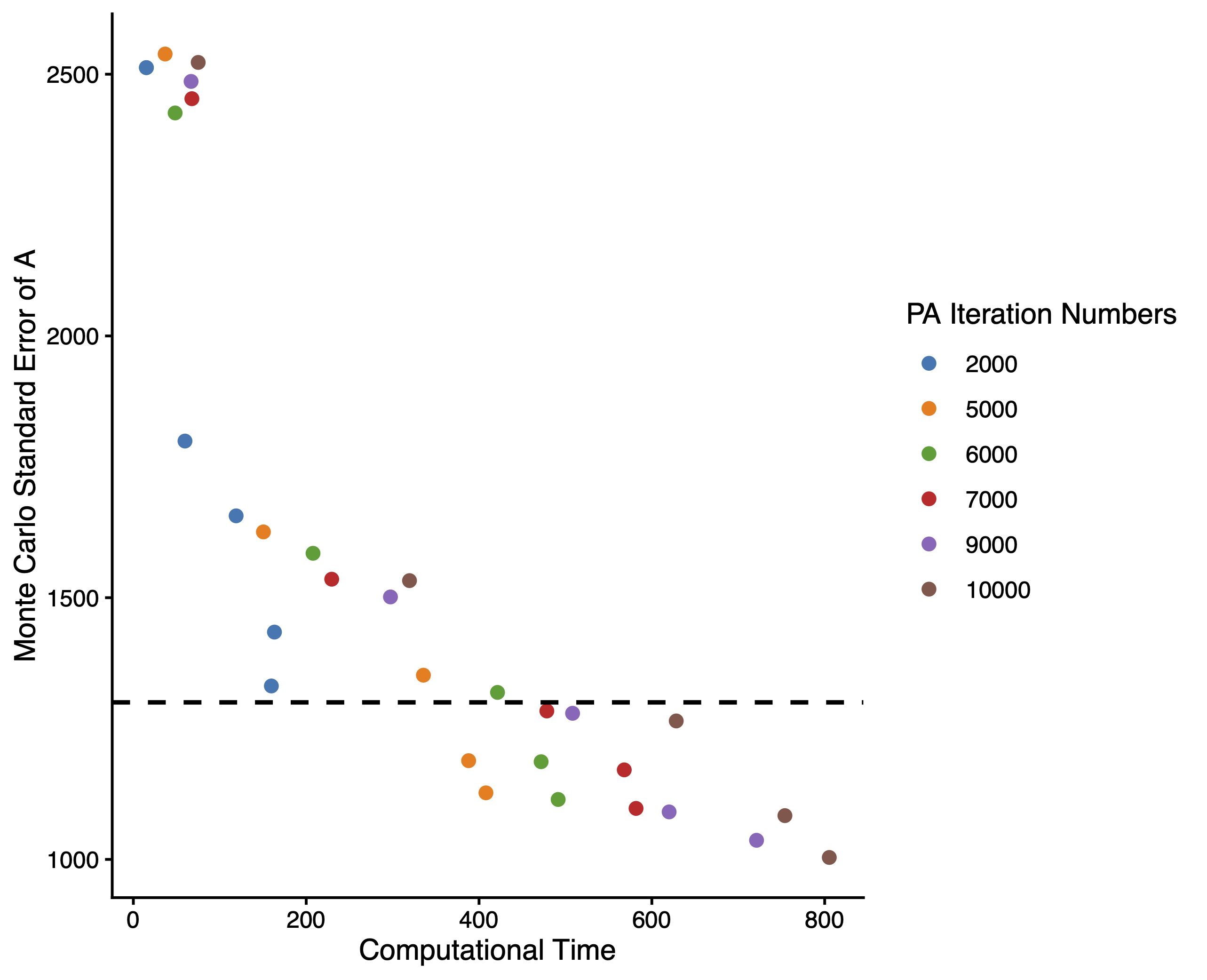} \\[0.5em]
\includegraphics[width=0.48\textwidth]{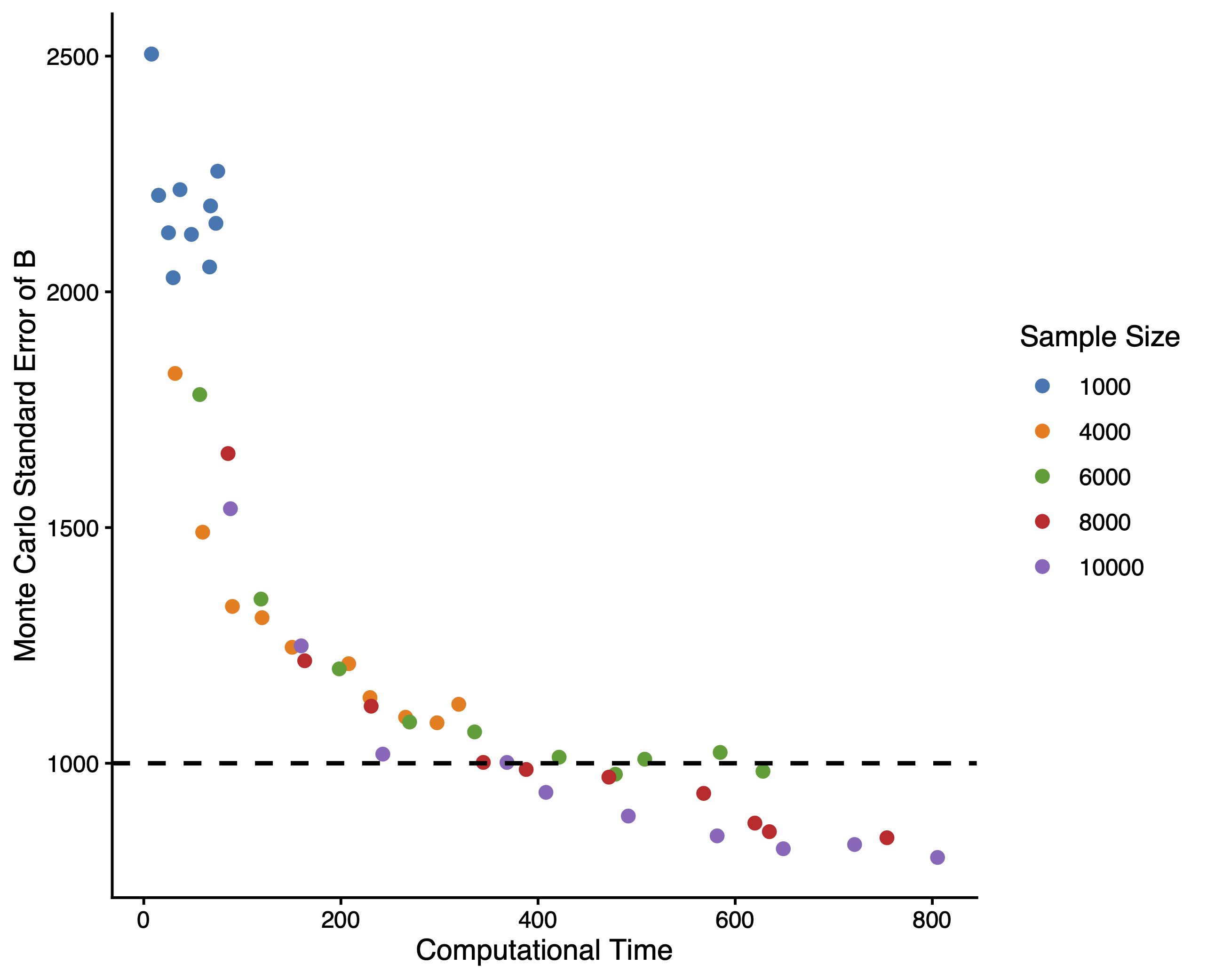} &
\includegraphics[width=0.48\textwidth]{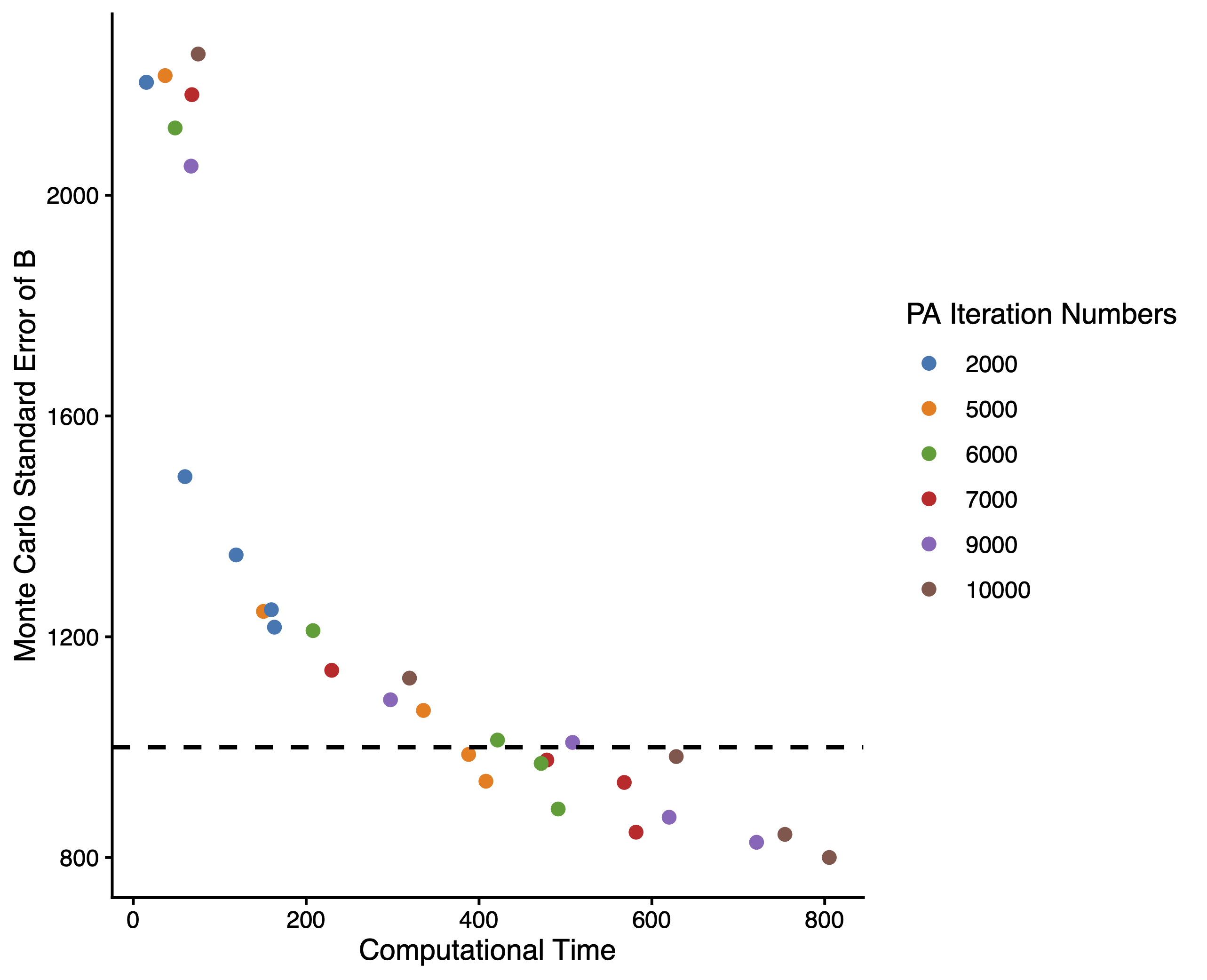} \\[0.5em]
\includegraphics[width=0.48\textwidth]{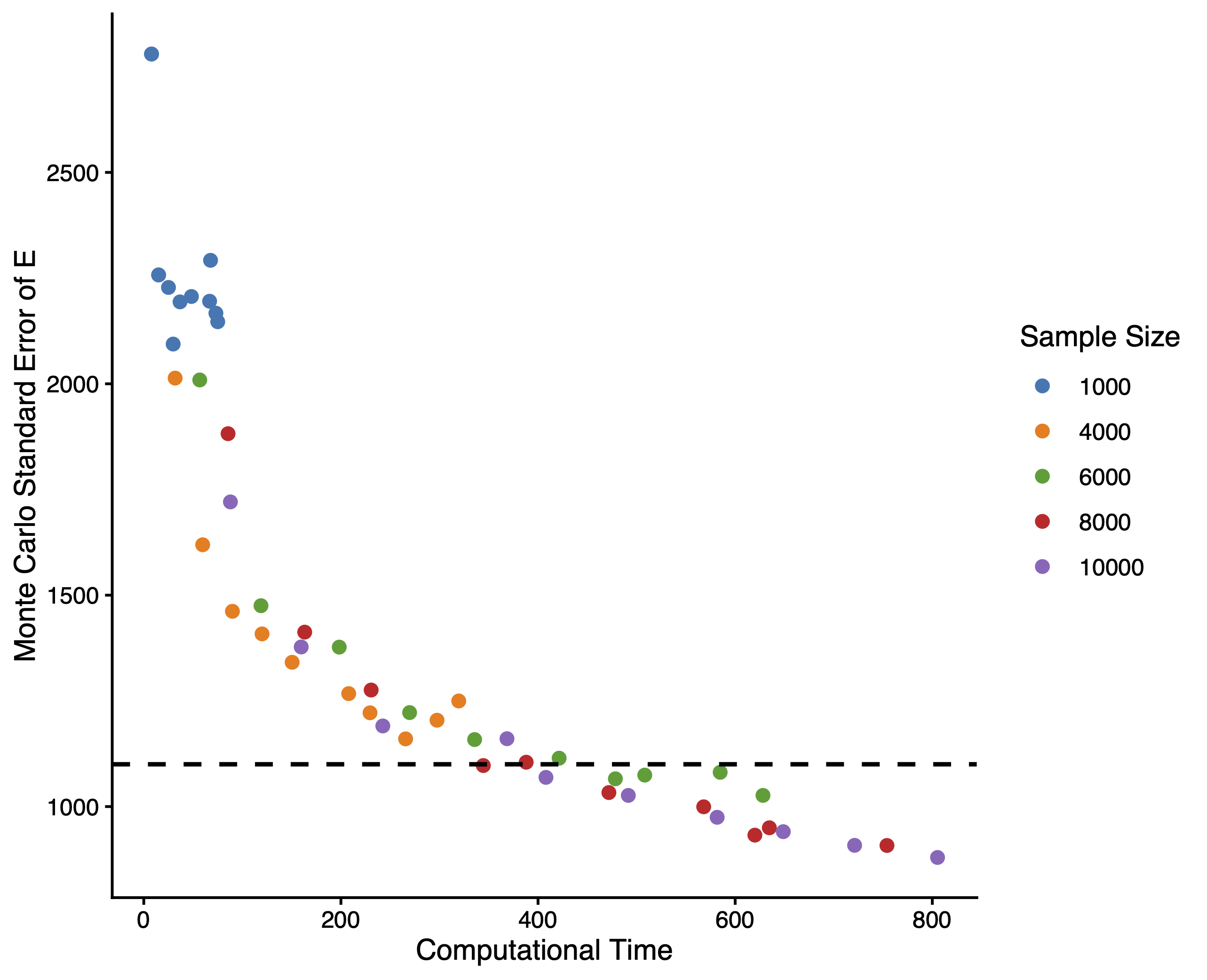} &
\includegraphics[width=0.48\textwidth]{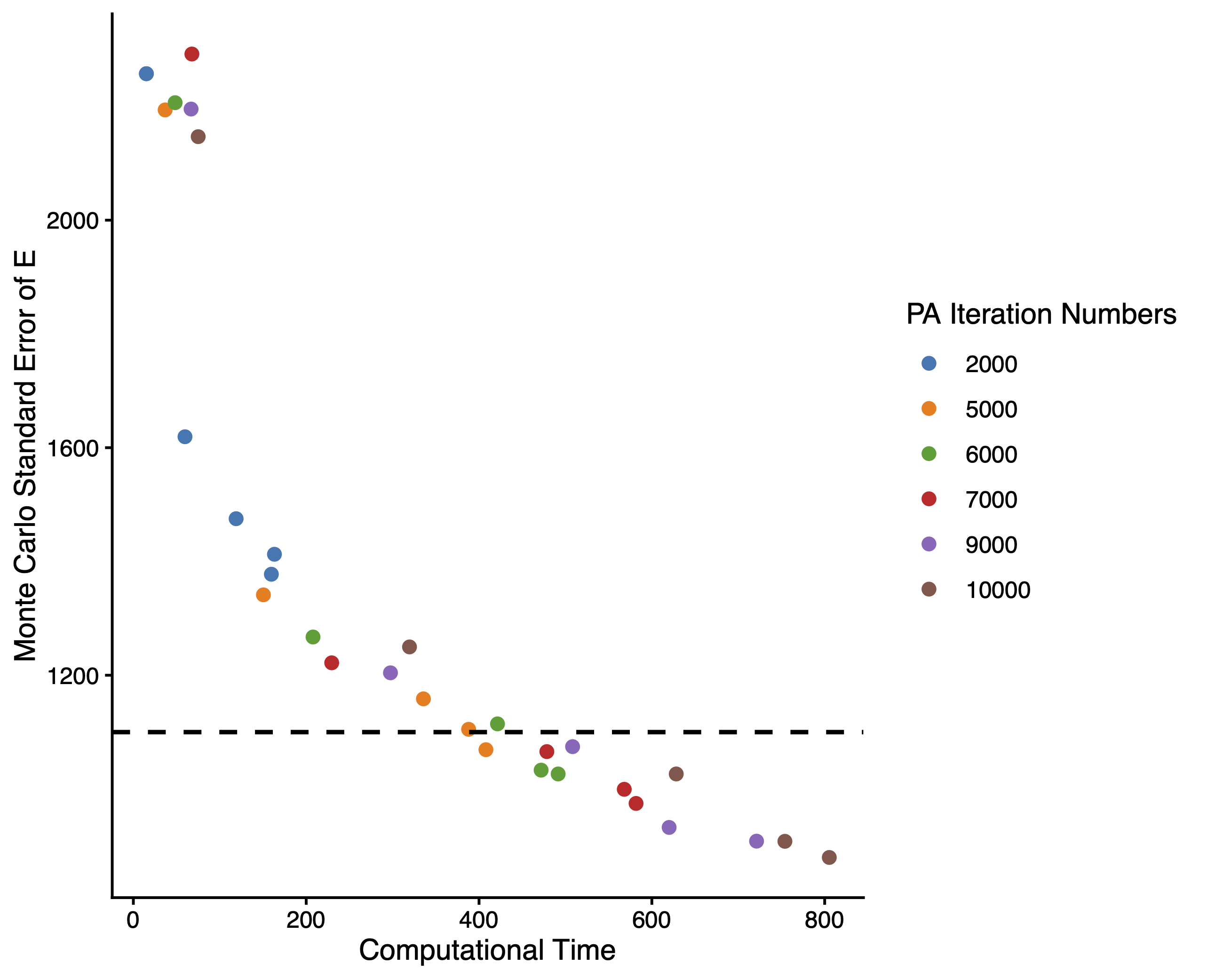} \\[0.5em]
\includegraphics[width=0.48\textwidth]{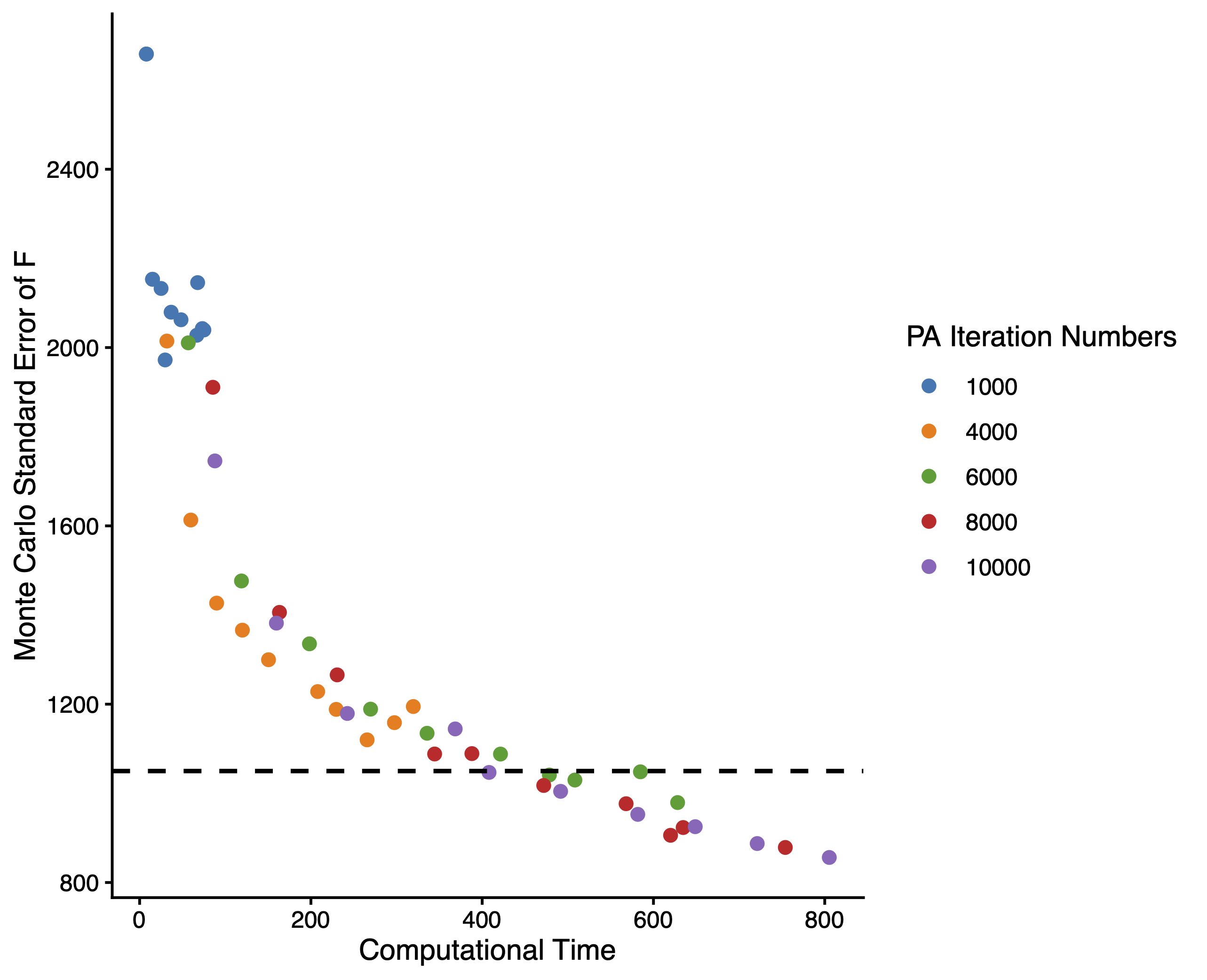} &
\includegraphics[width=0.48\textwidth]{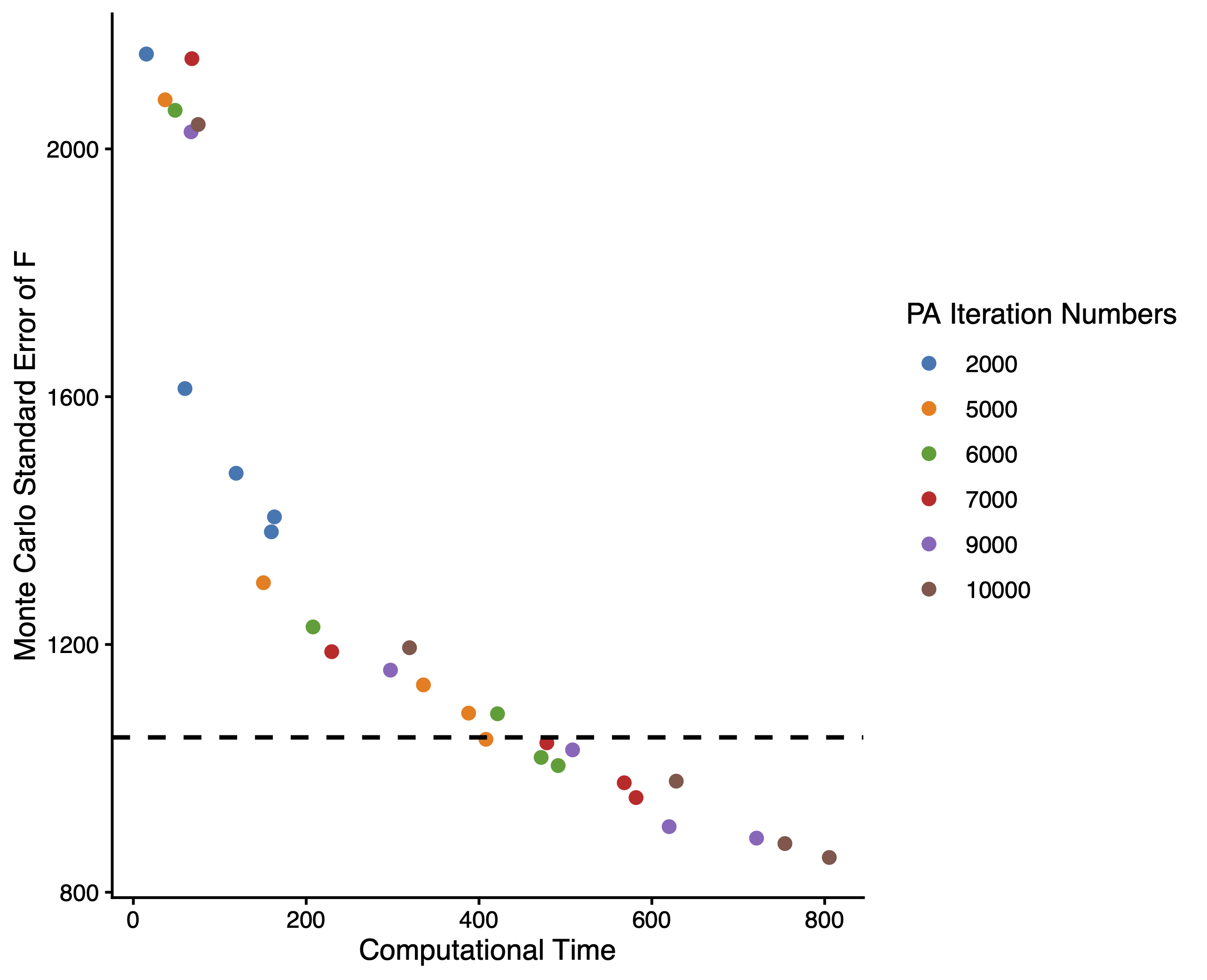} \\[0.5em]
\includegraphics[width=0.48\textwidth]{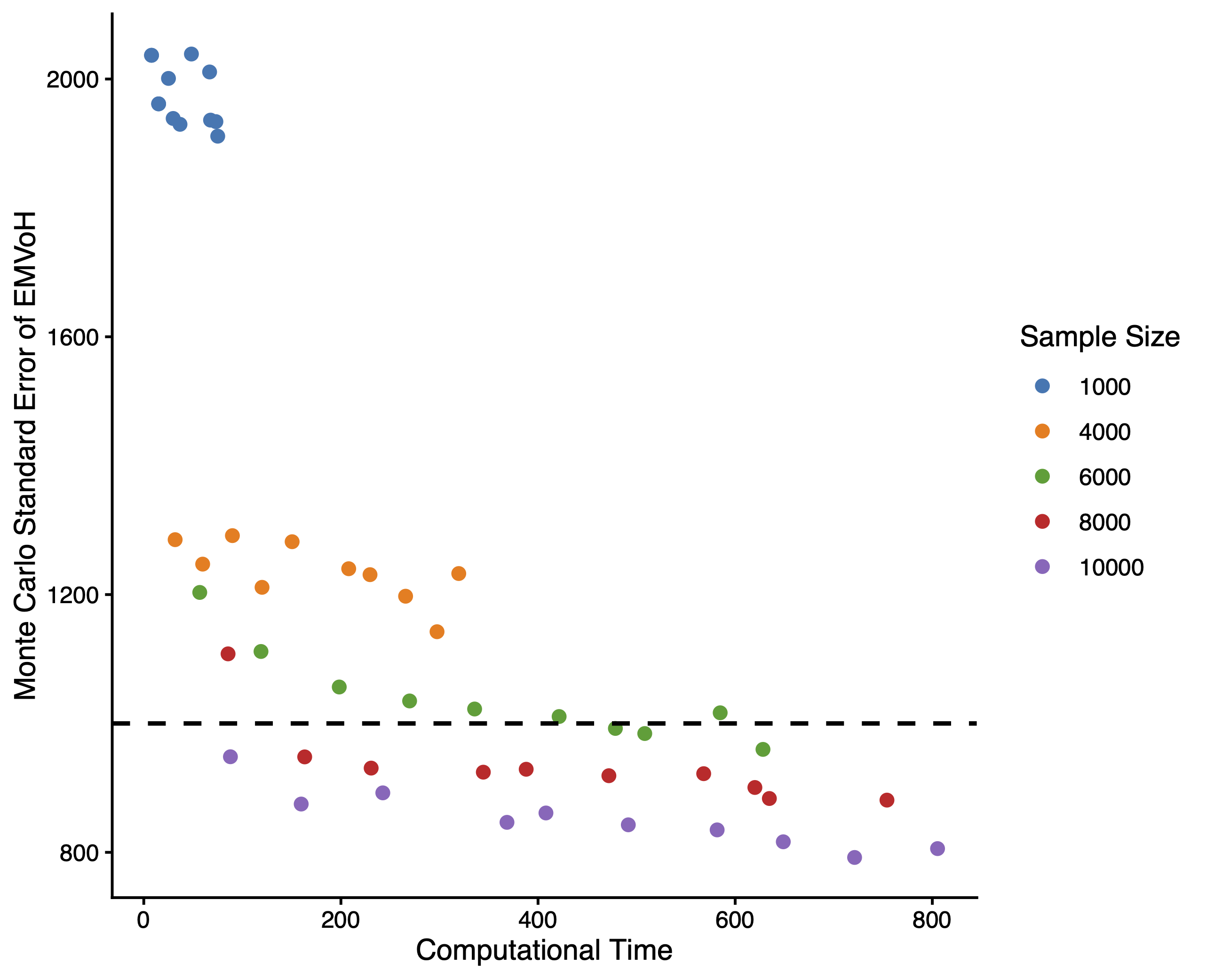} &
\includegraphics[width=0.48\textwidth]{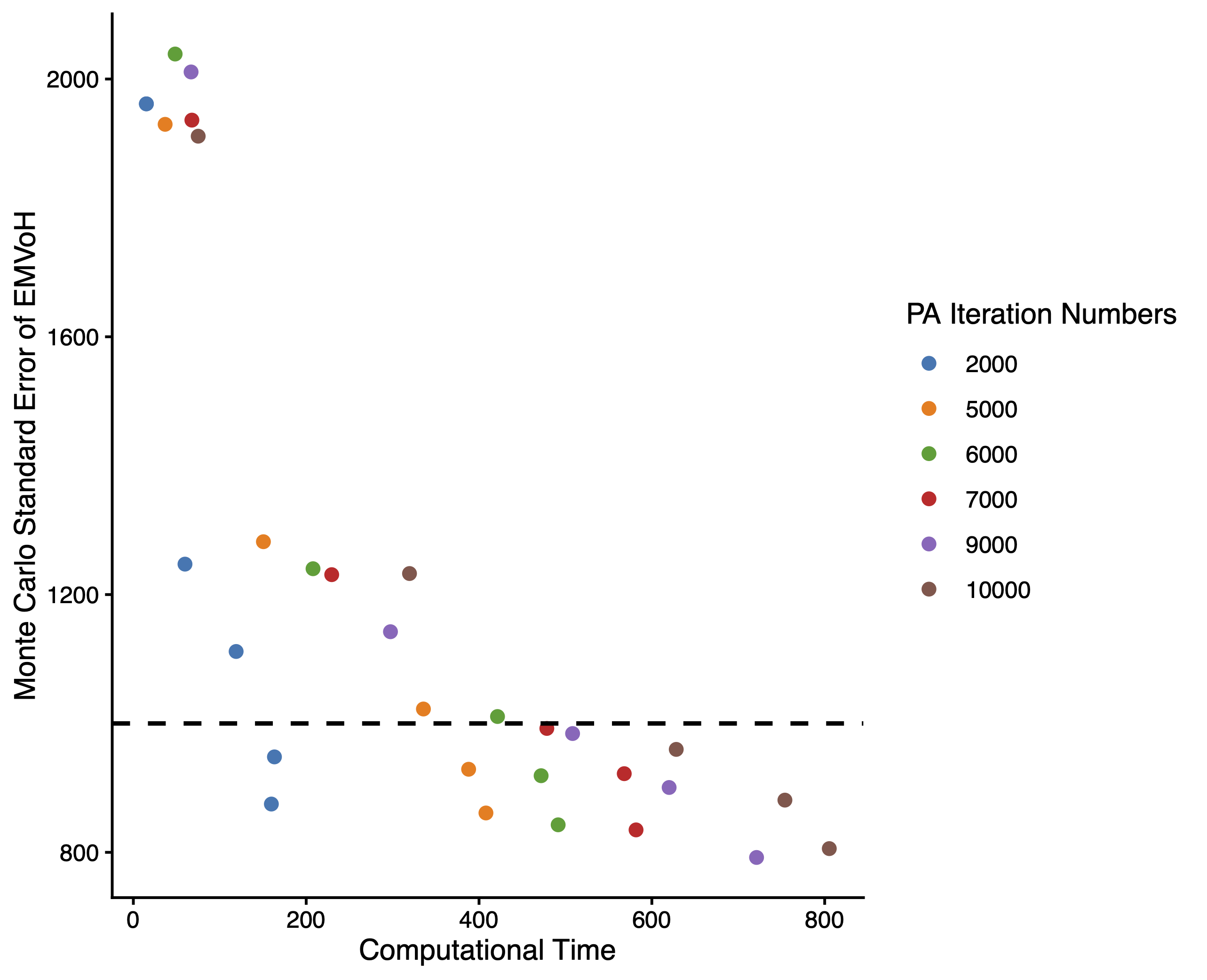} \\[0.5em]
\end{longtable}
\end{center}
\vspace{-4em}
\captionof{figure}{Computational time and the Monte Carlo standard errors of states $A$, $B$, $E$, $F$, and the expected maximum value of heterogeneity.}
\label{fig:time_mcse}

\newpage

\section*{Methods}
\subsection*{Boundaries for states $C$ and $D$}
The upper bound values for states $C$ and $D$ are states $E$ and $F$, respectively. The proof of the boundaries is shown in the following process.

From an individual-level costs and effectiveness perspective, the individual-level net benefit in the Sick-Sicker model can be calculated using the function: 
\begin{align} 
     nb_d &= \lambda \left[ \sum_{t=1}^{n_t} \Big(I\{ S_t = H \} \cdot e_H + I\{ S_t = S_1 \} \cdot e_{s_1} + I\{ S_t = S_2 \} \cdot e_{s_2}\Big) \times D_{e, t} \omega_t \right] -  \label{eq:my_equation} \\
     &\left[ \sum_{t=1}^{n_t} \Big(I\{ S_t = H \} \cdot c_H + I\{ S_t = S_1 \} \cdot c_{s_1} + I\{ S_t = S_2 \} \cdot c_{s_2} \Big) \times D_{c, t} \omega_t + I\{ S_t = D \} \cdot c_D \right], \notag 
\end{align} 
where $e_{\text{state}}$ and $c_{\text{state}}$ are the effectiveness and costs for each individual under the state at each time cycle, respectively. The components used to calculate $nb$ are referred to as $\boldsymbol{\mathcal{O}}$. $\boldsymbol{\mathcal{O}}$ can be divided into subsets of outcome components that are of interest and assumed to be perfectly predictable ($\boldsymbol{o}$) and remaining outcome components ($\boldsymbol{\mathcal{O}}^-$). We assume that the parameter sets of interest include $\boldsymbol{o}$ and $\boldsymbol{\omega}$, where $\boldsymbol{o} = (\boldsymbol{\omega}, \boldsymbol{\omega}^c)$. 

\begin{align*}
    C(\boldsymbol{o}) & = \mathbb{E}_{\boldsymbol{o}} \Big[ \max_d \mathbb{E}_{\boldsymbol{\mathcal{O}}^- \mid \boldsymbol{o}} \big[ \mathbb{E}_{\boldsymbol{\theta} \mid (\boldsymbol{\mathcal{O}}^-, \boldsymbol{o})} [ nb_d (\boldsymbol{\boldsymbol{o}, \mathcal{O}}^-, \boldsymbol{\theta}) ] \big] \Big] \\ 
    &= \mathbb{E}_{\boldsymbol{\omega}} \Big[ \mathbb{E}_{\boldsymbol{\omega}^c \mid \boldsymbol{\omega}} \big[ \max_d \mathbb{E}_{ (\boldsymbol{\mathcal{O}}^-, \boldsymbol{\theta}) \mid (\boldsymbol{\omega}, \boldsymbol{\omega}^c)} [ nb_d (\boldsymbol{\boldsymbol{w}, \boldsymbol{\omega}^c, \mathcal{O}}^-, \boldsymbol{\theta}) ] \big] \Big] \\ 
    &\geq \mathbb{E}_{\boldsymbol{\omega}} \Big[ \max_d \mathbb{E}_{\boldsymbol{\omega}^c \mid \boldsymbol{\omega}} \big[ \mathbb{E}_{ (\boldsymbol{\mathcal{O}}^-, \boldsymbol{\theta}) \mid (\boldsymbol{\omega}, \boldsymbol{\omega}^c)} [ nb_d (\boldsymbol{w}, \boldsymbol{\omega}^c, \boldsymbol{\mathcal{O}}^-, \boldsymbol{\theta}) ] \big] \Big] \\ 
    &= \mathbb{E}_{\boldsymbol{\omega}} \Big[ \max_d \mathbb{E}_{ (\boldsymbol{\theta}, \boldsymbol{\mathcal{O}}^-, \boldsymbol{\omega}^c) \mid \boldsymbol{\omega} } \big[ nb_d (\boldsymbol{\boldsymbol{w}, \boldsymbol{\omega}^c, \mathcal{O}}^-, \boldsymbol{\theta}) \big] \Big] \\
    &= C(\boldsymbol{\omega})
\end{align*}
As a result, the upper bound value of state $C$ equals the value of State $E$, which is the expected value of providing patients with individualized optimal intervention under current information. 

Similarly, the boundary of the state $D$ value is calculated through the following process:
\begin{align*}
    D(\boldsymbol{o}) &= \mathbb{E}_{\boldsymbol{\theta}} \Big[ \mathbb{E}_{\boldsymbol{o} \mid \boldsymbol{\theta}} \big[ \max_d \mathbb{E}_{\boldsymbol{\mathcal{O}}^- \mid (\boldsymbol{o}, \boldsymbol{\theta})} [ nb_d (\boldsymbol{\boldsymbol{o}, \mathcal{O}}^-, \boldsymbol{\theta}) ] \big] \Big] \\ 
    & = \mathbb{E}_{\boldsymbol{\theta}} \Bigg[ \mathbb{E}_{\boldsymbol{\omega} \mid \boldsymbol{\theta}} \Big[ \mathbb{E}_{\boldsymbol{\omega}^c \mid \boldsymbol{\omega}} \big[ \max_d \mathbb{E}_{\boldsymbol{\mathcal{O}}^- \mid (\boldsymbol{\omega}, \boldsymbol{\omega}^c, \boldsymbol{\theta}) } [ nb_d (\boldsymbol{\omega}, \boldsymbol{\omega}^c, \boldsymbol{\mathcal{O}}^-, \boldsymbol{\theta}) ] \big] \Big] \Bigg] \\
    & \geq \mathbb{E}_{\boldsymbol{\theta}} \Bigg[ \mathbb{E}_{\boldsymbol{\omega} \mid \boldsymbol{\theta}} \Big[ \max_d \mathbb{E}_{\boldsymbol{\omega}^c \mid \boldsymbol{\omega}} \big[ \mathbb{E}_{\boldsymbol{\mathcal{O}}^- \mid (\boldsymbol{\omega}, \boldsymbol{\omega}^c, \boldsymbol{\theta}) } [ nb_d (\boldsymbol{\omega}, \boldsymbol{\omega}^c, \boldsymbol{\mathcal{O}}^-, \boldsymbol{\theta}) ] \big] \Big] \Bigg] \\
    & = \mathbb{E}_{\boldsymbol{\theta}} \Big[ \mathbb{E}_{\boldsymbol{\omega} \mid \boldsymbol{\theta}} \big[ \max_d \mathbb{E}_{\boldsymbol{(\mathcal{O}}^-, \boldsymbol{\omega}^c) \mid (\boldsymbol{\omega}, \boldsymbol{\theta})} [ nb_d(\boldsymbol{\omega}, \boldsymbol{\omega}^c, \boldsymbol{\mathcal{O}}^-, \boldsymbol{\theta}) ]\big] \Big] \\
    & = D(\boldsymbol{\omega})
\end{align*}
As a result, the maximum value of state $D$ equals the value of State $F$, which is the value of providing patients with individualized optimal intervention under perfect parameter information.

\subsection*{Monte Carlo Standard Error}
We estimated the standard errors for the states $A$, $B$, $E$, and $F$ values in the framework from within-framework variance. Suppose the individuals included in the microsimulation models are independent, and the parameter sets used in the PA are independent. The formulas for the relationship between the variances and the number of individuals included in the microsimulation model and in the probabilistic analysis (PA) iterations are derived as follows.
\begin{align*}
    \text{Var}(A) & = \text{Var} \Big( \max_d \mathbb{E}_{\boldsymbol{\mathcal{O}}} \Big[ \mathbb{E}_{\boldsymbol{\theta} \mid \boldsymbol{\mathcal{O}} } [ nb_{ij} (d) ] \Big] \Big) \\
    & = \text{Var} \Big( \max \Big\{ \mathbb{E}_{\boldsymbol{\mathcal{O}}} \Big[ \mathbb{E}_{\boldsymbol{\theta} \mid \boldsymbol{\mathcal{O}} } [ nb_{ij} (d=1) ] \Big] , \dots, \mathbb{E}_{\boldsymbol{\mathcal{O}}} \Big[ \mathbb{E}_{\boldsymbol{\theta} \mid \boldsymbol{\mathcal{O}} } [ nb_{ij} (d={\mathcal{D}}) ] \Big]  \Big\} \Big) \\
    & \leq \text{Var} \Big(\mathbb{E}_{\boldsymbol{\mathcal{O}}} \Big[ \mathbb{E}_{\boldsymbol{\theta} \mid \boldsymbol{\mathcal{O}} } [ nb_{ij} (d=1) ] \Big] \Big) + \dots + \text{Var} \Big(\mathbb{E}_{\boldsymbol{\mathcal{O}}} \Big[ \mathbb{E}_{\boldsymbol{\theta} \mid \boldsymbol{\mathcal{O}} } [ nb_{ij} (d={\mathcal{D}}) ] \Big] \Big) \\
    & \approx \text{Var} \left( \frac{1}{n_i} \sum_{i = 1}^{n_i} \mathbb{E}_{\boldsymbol{\theta} \mid \boldsymbol{\mathcal{O}} } [ nb_{ij} (d=1) ] \right) + \dots + \text{Var} \left( \frac{1}{n_i} \sum_{i = 1}^{n_i} \mathbb{E}_{\boldsymbol{\theta} \mid \boldsymbol{\mathcal{O}} } [ nb_{ij} (d={\mathcal{D}}) ] \right) \\
    & \approx \frac{1}{n_i^2} \sum_{i = 1}^{n_i} \left[ \text{Var} \left( \frac{1}{n_{\text{PA}}} \sum_{j=1}^{n_{\text{PA}}} nb_{ij} (d=1) \right) + \dots + \text{Var} \left( \frac{1}{n_{\text{PA}}} \sum_{j=1}^{n_{\text{PA}}} nb_{ij} (d={\mathcal{D}}) \right) \right] \\
    & = \left( \frac{1}{n_i \cdot n_{\text{PA}}} \right)^2 \sum_{i = 1}^{n_i} \sum_{j=1}^{n_{\text{PA}}} \Big[ \sum_{d=1}^{\mathcal{D}} \text{Var}(\text{NB}_{ij} (d)) \Big]\\
\end{align*}
\vspace{-1cm}
\begin{align*}
    \text{Var}(B) & = \text{Var} \Big( \mathbb{E}_{\boldsymbol{\theta}} \Big[ \max_d \mathbb{E}_{\boldsymbol{\mathcal{O}} \mid \boldsymbol{\theta}} [nb_{ij} (d)] \Big] \Big) \\
    & \approx \text{Var} \left( \frac{1}{n_{\text{PA}}} \sum_{j=1}^{n_{\text{PA}}} \Big[ \max_d \mathbb{E}_{\boldsymbol{\mathcal{O}} \mid \boldsymbol{\theta}} \big[ nb_{ij} (d) \big] \Big] \right) \\
    & = \frac{1}{n_{\text{PA}}^2} \sum_{j=1}^{n_{\text{PA}}} \text{Var} \Big( \max_d \mathbb{E}_{\boldsymbol{\mathcal{O}} \mid \boldsymbol{\theta}} \big[ nb_{ij} (d) \big] \Big) \\
    & \leq \frac{1}{n_{\text{PA}}^2} \sum_{j=1}^{n_{\text{PA}}} \Bigg( \text{Var} \Big( \mathbb{E}_{\boldsymbol{\mathcal{O}} \mid \boldsymbol{\theta}} \big[ nb_{ij} (d = 1) \big] \Big) + \cdots + \text{Var} \Big( \mathbb{E}_{\boldsymbol{\mathcal{O}} \mid \boldsymbol{\theta}} \big[ nb_{ij} (d = {\mathcal{D}}) \big] \Big) \Bigg) \\ 
    & \approx \frac{1}{n_{\text{PA}}^2} \sum_{j=1}^{n_{\text{PA}}} \Bigg( \text{Var} \left( \frac{1}{n_i} \sum_{i=1}^{n_i} nb_{ij} (d = 1) \right) + \cdots + \text{Var} \left( \frac{1}{n_i} \sum_{i=1}^{n_i} nb_{ij} (d = {\mathcal{D}}) \right) \Bigg) \\
    & = \left( \frac{1}{n_{\text{PA}} \cdot n_i} \right)^2 \sum_{j=1}^{n_{\text{PA}}} \sum_{i = 1}^{n_i} \Big[ \sum_{d=1}^{\mathcal{D}} \text{Var}(nb_{ij} (d)) \Big]
\end{align*}
\begin{align*}
    \text{Var}(E) & = \text{Var} \Big( \mathbb{E}_{\boldsymbol{\mathcal{O}}} \Big[ \max_d \mathbb{E}_{\boldsymbol{\theta} \mid \boldsymbol{\mathcal{O}}} \big[ nb_{ij} (d) \big] \Big] \Big) \\ 
    & \approx \text{Var} \left( \frac{1}{n_i} \sum_{i=1}^{n_i} \max_d \mathbb{E}_{\boldsymbol{\theta} \mid \boldsymbol{\mathcal{O}}} \big[ nb_{ij} (d ) \big] \right) \\
    & = \frac{1}{n_i^2} \sum_{i=1}^{n_i} \text{Var} \Big( \max_d \mathbb{E}_{\boldsymbol{\theta} \mid \boldsymbol{\mathcal{O}}} \big[ nb_{ij} (d) \big] \Big) \\
    & \leq \frac{1}{n_i^2} \sum_{i=1}^{n_i} \Bigg( \text{Var} \Big( \mathbb{E}_{\boldsymbol{\theta} \mid \boldsymbol{\mathcal{O}}} \big[ nb_{ij} (d = 1) \big] \Big) + \cdots + \text{Var} \Big( \mathbb{E}_{\boldsymbol{\theta} \mid \boldsymbol{\mathcal{O}}} \big[ nb_{ij} (d = {\mathcal{D}}) \big] \Big) \Bigg) \\
    & \approx \frac{1}{n_i^2} \sum_{i=1}^{n_i} \Bigg( \text{Var} \left( \frac{1}{n_{\text{PA}}} \sum_{j = 1}^{n_{\text{PA}}} nb_{ij} (d = 1) \right) + \cdots + \text{Var} \left(\frac{1}{n_{\text{PA}}} \sum_{j = 1}^{n_{\text{PA}}} nb_{ij} (d={\mathcal{D}}) \right) \Bigg) \\
    &= \left( \frac{1}{n_i \cdot n_{\text{PA}}} \right)^2 \sum_{i = 1}^{n_i} \sum_{j=1}^{n_{\text{PA}}} \Big[ \sum_{d=1}^{\mathcal{D}} \text{Var}(nb_{ij} (d)) \Big]\\
\end{align*}
\begin{align*}
    \text{Var}(F) & = \text{Var} \Big( \mathbb{E}_{\boldsymbol{\mathcal{O}}} \Big[ \mathbb{E}_{\boldsymbol{\theta} \mid \boldsymbol{\mathcal{O}} } [ \max_d nb_{ij} (d) ] \Big] \Big) \\
    & \approx \frac{1}{n_i^2} \sum_{i=1}^{n_i} \text{Var} \Big( \mathbb{E}_{\boldsymbol{\theta} \mid \boldsymbol{\mathcal{O}} } \Big[ \max_d nb_{ij} (d) \Big] \Big) \\
    & \approx \left( \frac{1}{n_i \cdot n_{\text{PA}}} \right)^2 \sum_{i=1}^{n_i} \sum_{j=1}^{n_{\text{PA}}} \text{Var} \Big( \max_d nb_{ij} (d) \Big) \\
    & \leq \left( \frac{1}{n_i \cdot n_{\text{PA}}} \right)^2 \sum_{i=1}^{n_i} \sum_{j=1}^{n_{\text{PA}}} \Big[ \text{Var} \Big( nb_{ij} (d = 1) \Big) + \cdots + \text{Var} \Big( nb_{ij} (d={\mathcal{D}}) \Big) \Big] \\
    & \leq \left( \frac{1}{n_i \cdot n_{\text{PA}}} \right)^2 \sum_{i=1}^{n_i} \sum_{j=1}^{n_{\text{PA}}} \Big[ \sum_{d=1}^{\mathcal{D}} \text{Var} ( nb_{ij} (d) )  \Big]
\end{align*}

\subsection*{Estimation process for states $A$, $B$, $E$, and $F$ in the Sick-Sicker model example}
Figure~\ref{fig:calc_process} visualizes the estimation process for the values of states $A$, $B$, $E$, and $F$ under a specific correlation between the simulated individual-level net benefits, which provides the values of Expected Maximum Value of Heterogeneity (i.e., $F - A$), Expected Value of Perfect Information (i.e., $B - A$), and Maximum Value of Heterogeneity under current parameter information (i.e., $E - A$). Based on Eq.2, 5, and 6, the key point of the estimation process is to get the maximum expected net benefit.

Using the net benefit for each individual under each PA parameter set simulated from the decision-analytic model, we first specify the individualized optimal intervention. The individualized optimal intervention gives the highest net benefit for each individual under each PA iteration, $\max_d nb_d(\boldsymbol{\mathcal{O}}, \boldsymbol{\theta})$ (Figure~\ref{fig:calc_process}). By taking the expectation over the joint probability distribution of $\boldsymbol{\theta}$ using Monte Carlo simulations, the individualized optimal intervention under current parameter information can then be identified with the maximum value of expected net benefit for each individual ($\max_d \mathbb{E}_{\boldsymbol{\theta} \mid \boldsymbol{\mathcal{O}}} \big[ \text{NB}_d(\boldsymbol{\mathcal{O}}, \boldsymbol{\theta}) \big]$, Figure~\ref{fig:calc_process}). 

Similarly, the population-level optimal intervention under each parameter set can be determined by taking the expectation over the simulated outcome components $\boldsymbol{\mathcal{O}}$ and identifying the one with the maximum value of $\mathbb{E}_{\boldsymbol{\mathcal{O}} \mid \boldsymbol{\theta}} \big[ \text{NB}_d(\boldsymbol{\theta}) \big]$ (Figure~\ref{fig:calc_process}). 

{\centering
\includegraphics[width=0.9\textwidth]{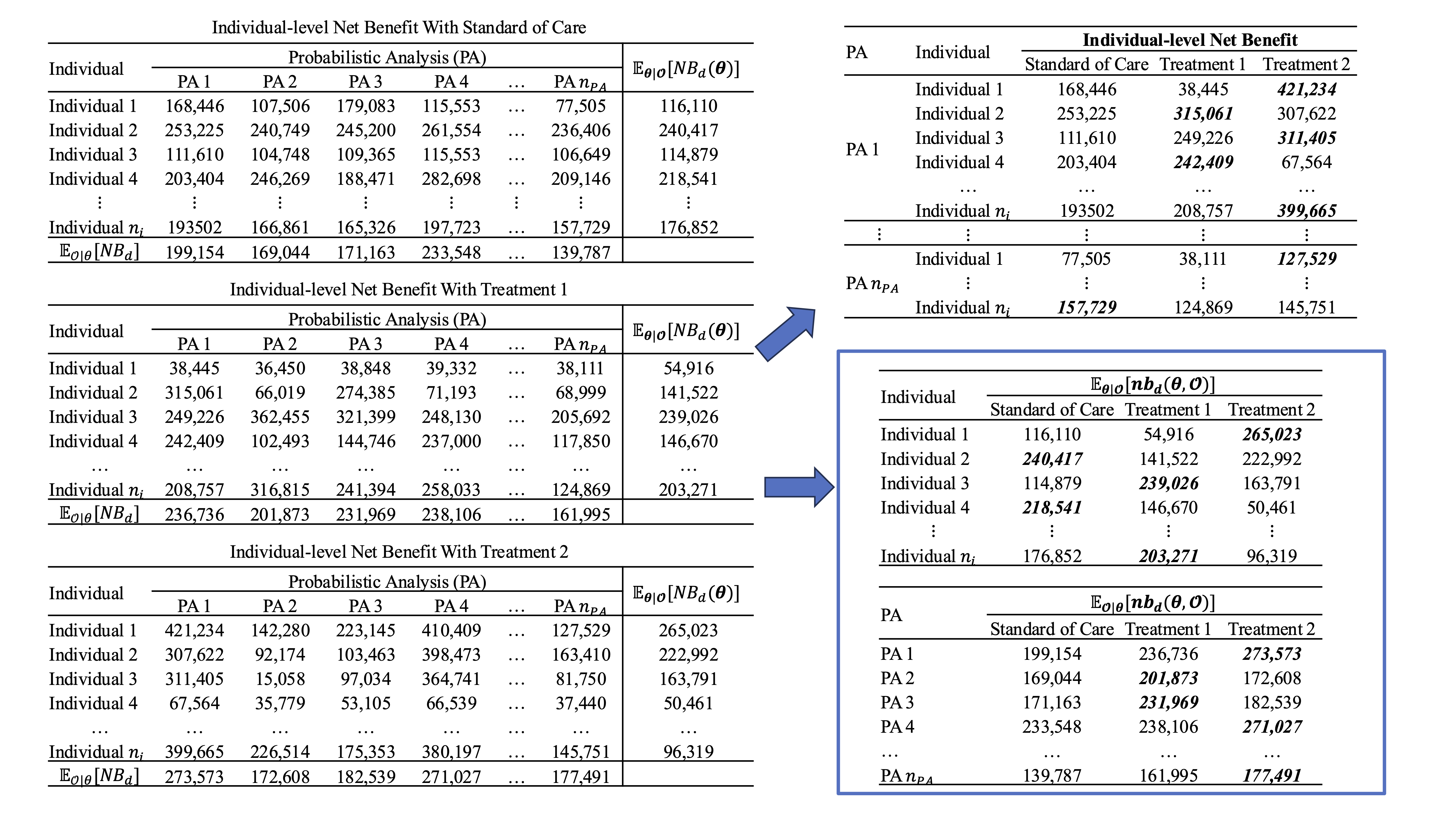}
\captionof{figure}{Estimate the values of research in the individualized optimal interventions using the Expected Value of Unexplained Heterogeneity framework. The process demonstrates how to calculate the $A$, $B$, $E$, and $F$ values included in the EVoUH framework. The individual net monetary benefit for each intervention and each probabilistic analysis parameter set is shown in the table.}
\label{fig:calc_process}
}

The values of states $A$, $B$, $E$, $F$, and the corresponding research directions can then be calculated using the listed maximum values.

\end{document}